\documentclass[twocolumn,english,superscriptaddress,pre,longbibliography]{revtex4-1}
\usepackage{xcolor}
\usepackage{bm}
\usepackage{amstext}
\usepackage{amssymb}
\usepackage{graphicx}
\usepackage{commath}
\usepackage{dsfont}
\usepackage{mathtools}
\usepackage{amsmath,mathrsfs}
\usepackage{physics}
\usepackage{hyperref}
\usepackage{algorithm}
\usepackage[end]{algpseudocode}
\pdfoutput=1
\algdef{SE}[DOWHILE]{Do}{doWhile}{\algorithmicdo}[1]{\algorithmicwhile\ #1}%
\hypersetup{colorlinks=true, linkcolor=blue, citecolor=blue, urlcolor=blue, unicode=true}

\begin{document}

\title{Large-scale portfolio optimization with variational neural annealing}

\author{Nishan Ranabhat}
\email{nranabha@uwaterloo.ca}
\affiliation{University of Waterloo, Waterloo, ON, Canada}
\affiliation{yiyaniQ, Toronto, ON, Canada}
\affiliation{Department of Physics, University of Maryland, Baltimore County, Baltimore, Maryland 21250, USA}

\author{Behnam Javanparast}
\affiliation{yiyaniQ, Toronto, ON, Canada}

\author{David Goerz}
\affiliation{Strategic Frontier Management, Dana Point, CA, USA}

\author{Estelle Inack}
\email{estelle@yiyaniq.com}
\affiliation{Perimeter Institute, Waterloo, ON, Canada}
\affiliation{yiyaniQ, Toronto, ON, Canada}
\affiliation{University of Waterloo, Waterloo, ON, Canada}

\date{\today}

\begin{abstract}
Portfolio optimization is a routine asset management operation conducted in financial institutions around the world. However, under real‐world constraints such as turnover limits and transaction costs, its formulation becomes a mixed‐integer nonlinear program that current mixed-integer optimizers often struggle to solve. We propose mapping this problem onto a classical Ising-like Hamiltonian and solving it with Variational Neural Annealing (VNA), via its classical formulation implemented using autoregressive neural networks. We demonstrate that VNA can identify near-optimal solutions for portfolios comprising more than 2,000 assets and yields performance comparable to that of state-of-the-art optimizers, such as Mosek, while exhibiting faster convergence on hard instances. Finally, we present a dynamical finite-size scaling analysis applied to the S\&P 500, Russell 1000, and Russell 3000 indices, revealing universal behavior and polynomial annealing time scaling of the VNA algorithm on portfolio optimization problems.

\end{abstract}
\maketitle

\section{Introduction} \label{sec:introduction}

The question of how to choose decision variables to minimize an objective, often referred to as an optimization problem, arises in many physical and non‐physical contexts. In physics, one encounters examples such as minimizing Lennard–Jones clusters~\cite{northby1987structure, doye1998thermodynamics} or finding ground states of Ising spin glasses~\cite{mezard1987spin, Spin_Glass}. Outside of physics, similar optimization problems appear in biology~\cite{Banga2008,RealiPriamiMarchetti2017,Naseri2020}, scheduling \cite{Abdalkareem2021,DAUZEREPERES2024409,XIONG2022105731}, and finance~\cite{markowitz1952}, to name a few. When the decision variables are discrete or a mixture of discrete and continuous, the problem becomes intractable (NP-hard in the worst case), and heuristic methods are required to find near-optimal solutions~\cite{barahona1982computational, lucas2014ising}.  Intractability can also stem from decision variable interactions, whether k-local couplings akin to k-SAT, or nonlinear constraints characteristic of mixed-integer nonlinear programs (MINLPs)~\cite{belotti2013mixed}. One prominent MINLP instance is the portfolio optimization (PO) problem. 

The Modern Portfolio Theory (MPT) frames portfolio optimization as allocating capital among a set of assets to maximize expected return subject to an investor’s risk tolerance, preferences, and business or regulatory constraints~\cite{markowitz1952}. In its simplest form, the MPT model maximizes risk-adjusted returns with a budget constraint.  However, enforcing discrete choices (e.g., including or excluding an asset) with cardinality constraints transforms PO into an NP-hard problem.  In fact, the binary PO formulation maps directly onto an Ising Hamiltonian (where each asset corresponds to a spin) and finding its ground state is known to be NP-complete~\cite{sherrington1975solvable, lucas2014ising}. Furthermore, nonlinear constraints can also introduce NP-hardness even when considering continuous decision variables. That is the case for PO with long-short positions and leverage constraints~\cite{galluccio1998rational}. The latter, when combined with transaction costs that necessitate buy or sell decisions, yields a MINLP that is typically addressed using advanced algorithms that utilize relaxation techniques in conjunction with branch-and-cut solvers. However, such approaches struggle to scale beyond a few hundred assets and may become trapped in local minima in the worst-case scenario.  Practical taxable PO—for instance, modeling multiple share lots per security with distinct capital-gains tax rates—yields another class of highly nonconvex MINLPs that dramatically increase computational time or render exact methods intractable~\cite{Moehle2021}.


To address these issues, researchers have explored quantum computing to tackle portfolio optimization, alongside quantum-inspired algorithms that run on classical hardware; the expectation being that quantum superposition and tunneling would lead to a faster and more efficient exploration of the solution space~\cite{orus2019quantum, egger2020quantum, herman2023quantum, rebentrost2024quantum,naik2025portfolio}. Quantum computing has mainly been implemented via quantum annealing~\cite{johnson2011quantum, rosenberg2015solving, venturelli2019reverse, Multiverse, palmer2021quantum} and gate-based quantum algorithms~\cite{slate2021quantum, brandhofer2022benchmarking,baker2022wasserstein, buonaiuto2023best}. Quantum annealing hardware directly programs the Ising Hamiltonian but faces challenges such as embedding overhead, sparse qubit connectivity, analog noise, and limited system size.  Gate-based quantum algorithms, such as the Quantum Approximate Optimization Algorithm (QAOA)~\cite{farhi2014quantum} or the Variational Quantum Eigensolver (VQE)~\cite{peruzzo2014variational}, encode the portfolio objective into a parameterized circuit.  These methods encounter difficulties with circuit depth requirements, noise, decoherence, barren plateaus, and Hamiltonian encoding overhead. Strategies to mitigate these challenges, such as error mitigation techniques, adaptive circuit design, problem-specific mixers, and counterdiabaticity, are actively being investigated~\cite{blekos2024review, abbas2024challenges, simen2025branch}. 

Quantum-inspired methods such as tensor networks~\cite{schollwock2011density, Multiverse}, simulated annealing~\cite{SA, Lang2022}, digital annealing~\cite{matsubara2018ising, aramon2019physics}, and simulated bifurcation machines~\cite{steinhauer2020solving} have also been explored. Tensor networks have successfully simulated PO problems with hundreds of assets. Although the required bond dimension grows rapidly with system size, incurring substantial classical memory and computational costs~\cite {Multiverse}. Simulated annealing and digital annealing leverage classical stochastic processes or specialized hardware to find near-optimal solutions, typically outperforming existing quantum hardware on practical problem sizes~\cite{aramon2019physics, Lang2022} though having efficiency issues in glassy dynamics due to the long autocorrelation times. Simulated bifurcation machines, inspired by nonlinear dynamical systems, have demonstrated the ability to rapidly approximate optimal portfolios even for hundreds of assets, showcasing promising scalability~\cite{tatsumura2024}. However, they are expected to struggle in handling the full complexity of nonconvex nonlinear mixed-integer portfolio optimization at scale because of the sheer complexity of encoding those constraints into a Quadratic Unconstrained Binary Optimization (QUBO) problem. Most of these quantum and quantum-inspired techniques require reformulating the PO problem in the form of a QUBO problem which is a constraint that can be circumvented using machine learning approaches. 


Machine learning research has recently seen great interest in tackling the portfolio optimization problem~\cite{Alcazar2024, Zhang_2020, LopezDePrado2024, Mienye2024, Li2024, Cao2024}. In that setting, the objective function is viewed as a loss function or a reward function that needs to be optimized either through pure variational methods~\cite{Zhang_2020} or using reinforcement learning methods~\cite{Sood2023, Jeon2024}. Different neural networks have been used to generate the decision variables used to train the objective function. They range from feed-forward neural networks~\cite{Zhang_2020} to graph neural networks~\cite{Ekmekcioglu2023} to recurrent neural networks~\cite{Cao2024} and transformers~\cite{Mai_2024}. End-to-end simulations~\cite{Wei2023, Ndikum2024, Uysal2024} that do not require using the traditional MPT formulation have also been introduced recently, showing superior performance compared to their mean-variance counterparts~\cite{Sood2023, Uysal2024}. These machine learning approaches are equivalent to the variational optimization of a classical Hamiltonian representing the financial portfolio objective function, which has been shown to converge to local minima~\cite{VNA, khandoker2025latticeproteinfoldingvariational}. Furthermore, they are mainly focused on solving exchange-traded funds (ETFs)~\cite{Zhang_2020, Uysal2024} or asset pricing problems~\cite{Mai_2024, cao2025deeplearningllmssurvey} rather than equity portfolios, albeit for a small universe of stocks~\cite{Deng2024}. 

In this article, we augment variational optimization with the annealing paradigm using the variational neural annealing (VNA) algorithm~\cite{VNA}. Previously, it was successfully applied to disordered Ising glasses~\cite{biazzo2024sparse, ma2024message, liu2025efficient}, combinatorial optimization problems~\cite{khandoker2023supplementing}, quantum matter simulations~\cite{hibat2023investigating, hibatallah2024recurrentneuralnetworkwave, moss2025leveragingrecurrenceneuralnetwork}, and lattice protein folding~\cite{khandoker2025latticeproteinfoldingvariational}. Here, we implement VNA to solve large-scale optimization problems for equity portfolios. We utilize recurrent neural networks as variational ansatzes, aligned with the classical formulation of VNA due to its more favorable computational complexity scaling. This approach enables simulations of large-scale financial portfolios containing more than 2000 assets that are relevant to solving today's real-world financial optimization problems. We primarily benchmark our algorithms on a MINLP using real financial data and demonstrate their ability to find valid portfolios comparable to commercial solvers such as Mosek~\cite{mosek}. We also perform a dynamical finite-size scaling study on the VNA algorithm and demonstrate a universal scaling behavior of VNA trained across different trading periods on the S\&P 500, Russell 1000, and Russell 3000 indexes. 

The remainder of this article is organized as follows. In Sec.~\ref{sec:po}, we present the portfolio optimization problem and its MINLP formulation. In Sect.~\ref{sec:methods} we review the Variational Neural Annealing method implemented with recurrent neural networks. In Sect.~\ref{sec:resutls} we report VNA benchmarks on the S\&P 500, compare large-scale portfolio results against Mosek, and report on dynamical finite-size scaling studies. We conclude in Sect.~\ref{sec:conclusions}.
\section{Portfolio Optimization Formulation} \label{sec:po}

\begin{figure*}[t!]
    \centering
    \includegraphics[width=1.0\linewidth]{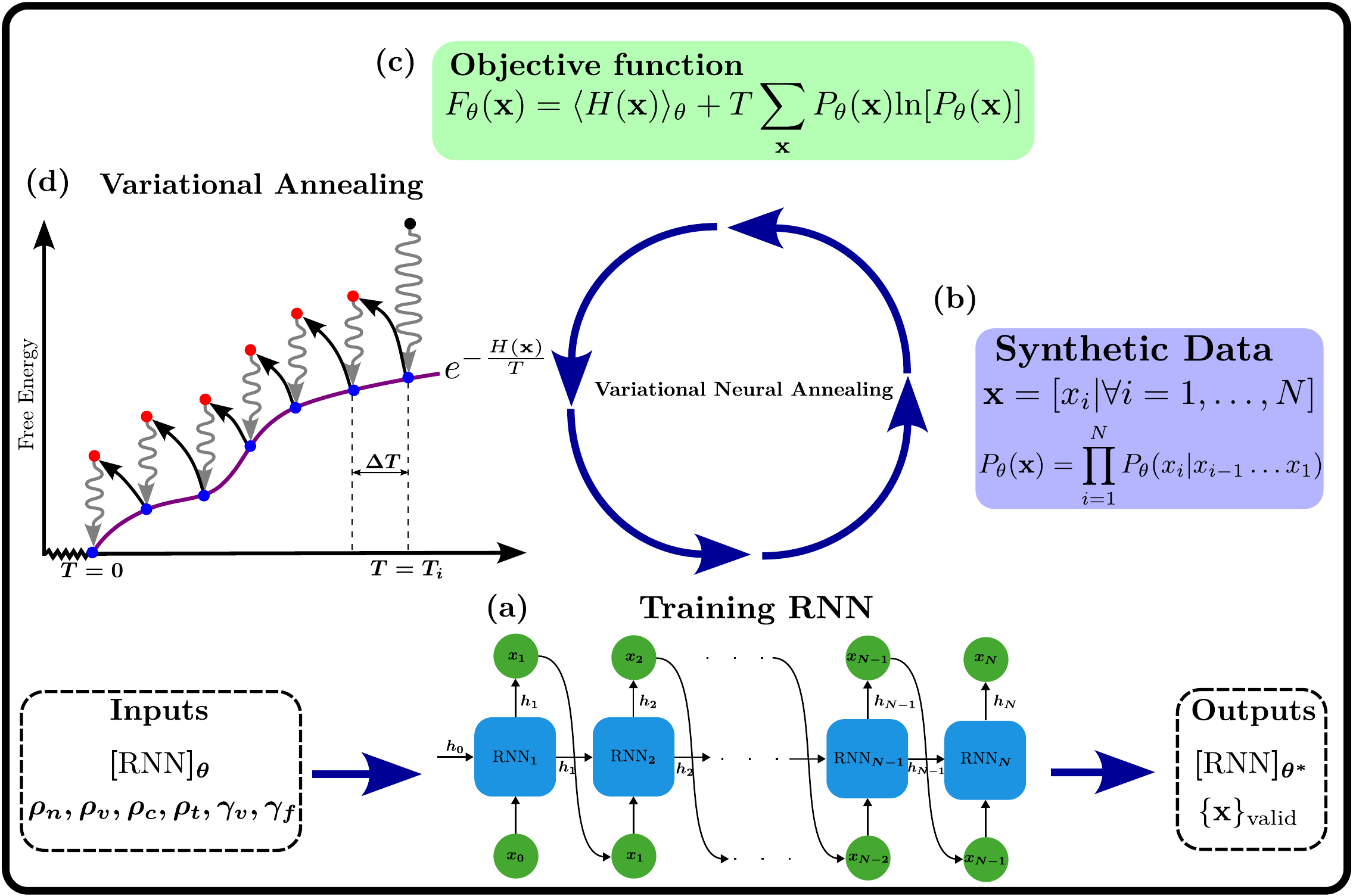}
    \caption{Variational Neural Annealing (VNA) algorithm for portfolio optimization. (a) A recurrent neural network (RNN) is employed as an expressive variational ansatz to learn the probability distribution of the portfolio weights $P(\mathbf{x})$. It takes the model parameters and the parameters of the portfolio optimization Hamiltonian as input and returns the optimal model and valid solutions as output. (b) The synthetic data $\{\mathbf{x}, P(\mathbf{x})\}$ generated by RNN. (c) Calculating the objective function (the variational free energy) $F_{\theta}(\mathbf{x})$. (d) Variational annealing: starting at $T_i \gg 0$, RNN is trained to equilibrate to the Boltzmann distribution $\sim e^{-\frac{H(\mathbf{x})}{T}}$ at every temperature until $T=0$.}
    \label{fig:VNA_protocol}
\end{figure*}

The Modern Portfolio Theory also known as the Markowitz mean-variance model~\cite{markowitz1952} focuses on maximizing the return on investment by building a diversified portfolio that minimizes risk within a specified threshold, achieved through the selection of an optimal allocation of portfolio weights. For a portfolio with $N$ assets, the vector $\mathbf{x} = [ x_i \hspace{0.1cm} | \hspace{0.1cm} \forall i = 1,\ldots,N]$ defines the fraction of the total investment allocated to each asset for long-only portfolios. The goal is to select optimal portfolio weights that maximize the return on investment while minimizing risk. Quantitatively, this can be set up as an optimization problem with the objective function defined as
\begin{equation}\label{eq:1}
    H(\mathbf{x}) = -\bm{\mu}^T \mathbf{x} + \rho_{\text{v}} \mathbf{x}^T\mathbf{C} \mathbf{x},
\end{equation}
where $\bm{\mu} = [\mu_i \hspace{0.1cm} | \hspace{0.1cm} \forall i = 1,\ldots,N]$ is the relative price return, which is calculated by averaging the historical returns. $\mathbf{C} = [C_{ij} \hspace{0.1cm} | \hspace{0.1cm} \forall i,j = 1,\ldots,N]$ is a covariance matrix that quantifies the correlation between different assets within the portfolio computed over a chosen look-back window. The factor $\rho_{\text{v}}$ is known as the risk aversion constant that regulates the risk penalty that an investor is willing to bear. In Eq.~(\ref{eq:1}), the first term quantifies the total expected return. In contrast, the second term which is portfolio variance quantifies the risk as a high degree of inter-asset correlation makes the portfolio more volatile and therefore more risky.

Optimization of the objective function in Eq.~(\ref{eq:1}) must be subject to a normalization constraint, which guarantees that the individual investment in each asset amounts to at most the total money invested. In our formulation, we impose full investment of the budget in the form of an equality constraint which is enforced by adding a penalty term to the objective function as

\begin{equation}\label{eq:2}
    H(\mathbf{x}) = -\bm{\mu}^T \mathbf{x} + \rho_{\text{v}}\, \mathbf{x}^T\mathbf{C} \mathbf{x} + \rho_{\text{n}} \left(\mathbf{1}^T \mathbf{x} - 1\right)^2,
\end{equation}
where $\mathbf{1}^T$ is a column vector of ones. A good portfolio selection maximizes the expected return while minimizing overall risk, which is quantified by the ratio of return to volatility known as the Sharpe ratio.
\begin{equation}\label{eq:3}
\text{Sharpe ratio} = \frac{\bm{\mu}^T \mathbf{x}}{\sqrt{\mathbf{x}^T\mathbf{C} \mathbf{x}}}.
\end{equation}
The covariance matrix $\mathbf{C}$ is positive semi-definite by construction, and the objective function is therefore convex. In this case, the objective function has a well-defined single global minimum and can be solved exactly \cite{Multiverse}. This is assuming $\mathbf{x} \in \mathbb{R}^N$, that is, the elements of vector $\mathbf{x}$ take real continuous values; however, in some financial settings, the elements of $\mathbf{x}$ take values in discrete packets. This converts the convex optimization problem into a discrete combinatorial optimization problem, akin to the fully connected disordered Ising Hamiltonian \cite{Spin_Glass}, and therefore PO becomes NP-hard.

\subsection{Transaction Costs}

The objective function defined in Eq.~(\ref{eq:2}) represents an idealized scenario. In real financial markets, transaction costs are incurred for every trade and are deducted from the total investment. The transaction cost can be of two types: the first is variable cost, which is proportional to the volume of the trade, and the second is fixed cost, which is proportional to the number of trades, independent of the trade volume. These costs are defined as
\begin{equation}\label{eq:4}
    \text{Cost}_{v}(\mathbf{x}_t,\mathbf{x}_{t-1}) = \gamma_v \sum_i |x_{i,t-1}-x_{i,t}|.
\end{equation}

\begin{equation}\label{eq:5}
    \text{Cost}_{f}(\mathbf{x}_t,\mathbf{x}_{t-1}) = \gamma_f \sum_i (1-\delta_{x_{i,t-1}}^{x_{i,t}}).
\end{equation}
Incorporating these costs into the objective function creates discontinuities and jumps, rendering brute-force gradient-based methods infeasible. Commercial solvers, such as Mosek, resort to computationally expensive mixed-integer programming (MIP) for optimization. The VNA technique that we employ is independent of the differentiability and convexity of the objective function. Therefore, the optimization strategy remains the same whether or not transaction costs are considered. We incorporate the transaction costs in the objective function in two ways: first, the normalization penalty term is altered to include the transaction costs, 
\begin{equation}\label{eq:6}
\rho_{\text{n}} (\mathbf{1}^T \mathbf{x} + \text{Cost}_{v} + \text{Cost}_{f} - 1)^2.
\end{equation}
Second, the total fixed transaction cost is limited by penalizing the fixed cost above the target fixed cost $C$,
\begin{equation}\label{eq:7}
\text{Penalty}_{\text{cost}} =
\begin{cases} 
    \rho_{c} (\text{Cost}_{f}- C)^2, & \text{Cost}_{f} > C \\
    0, & \text{otherwise}.
\end{cases}
\end{equation}

\subsection{Other Financial Constraints}

In real financial settings, several additional practical constraints arise beyond the transaction cost and normalization constraints. We incorporate two further constraints: the first is the volatility constraint, which penalizes volatility beyond a specified bound $V$. The volatility constraint is incorporated in the objective function as  
\begin{equation}\label{eq:8}
\text{Penalty}_{\text{vol}} =
\begin{cases} 
    \rho_{\text{v}} \Big(\sqrt{\mathbf{x}^T\mathbf{C} \mathbf{x}}- V\Big)^2, & \sqrt{\mathbf{x}^T\mathbf{C} \mathbf{x}} > V \\
    0, & \text{otherwise}.
\end{cases}
\end{equation}
The other constraint is the turnover constraint, which limits the number of trades within the rebalancing period, thus limiting the total transaction costs. This constraint is incorporated by penalizing the changes in portfolio weights with respect to a benchmark portfolio weight beyond a specified bound $T$,
\begin{equation}\label{eq:9}
\text{Penalty}_{\text{turn}} =
\begin{cases} 
    \rho_{\text{t}} \big(\sum_i (x_i - \tilde{x}_i)- T\big)^2, & \sum_i (x_i - \tilde{x}_i) > T \\
    0, & \text{otherwise}.
\end{cases}
\end{equation}
Penalties in Eqs.~(\ref{eq:7}), (\ref{eq:8}), and (\ref{eq:9}) are defined as piecewise functions that can be incorporated into the objective function by using the RELU function. 
\section{Methods} \label{sec:methods}

Traditionally, simulated annealing (SA) has been considered a standard method to approach non-convex optimization problems with a complex energy landscape \cite{SA} and has far-reaching implications~\cite{SA1,SA2,SA3,SA4}. However, SA gives an exact solution only in the limit of an infinitely slow annealing schedule; for any finite annealing schedule, the solution diverges significantly as SA fails to exactly sample the equilibrium Boltzmann distribution \cite{VNA}. This issue becomes more pronounced at low temperatures, where it is more challenging to generate independent samples using the Metropolis algorithm due to high autocorrelation. Recently, variational neural annealing has been proposed to address the limitations of SA which combines annealing with variational optimization of the autoregressive neural networks~\cite{VNA}. The primary difference between SA and VNA lies in the sampling of the Boltzmann distribution $\sim e^{-\frac{H(\mathbf{x})}{T}}$. While SA employs the Metropolis algorithm to sample the Boltzmann distribution at a given temperature, VNA implements an expressive parametrized variational ansatz to learn the probability distribution $P_{\theta}(\mathbf{x}) \rightarrow e^{-\frac{H(\mathbf{x})}{T}}$. This is achieved by minimizing the variational free energy at each annealing step,

\begin{align}\label{eq:10}
    \nabla_{\theta}F_{\theta}(\mathbf{x}) =0: F_{\theta}(\mathbf{x})  = \langle H(\mathbf{x}) \rangle_{\theta} + T \sum_{\mathbf{x}} P_{\theta}(\mathbf{x}) \text{ln}[P_{\theta}(\mathbf{x})],
\end{align}
where, $\theta$ represents all the learnable variational parameters of the ansatz, $\langle H(\mathbf{x}) \rangle_{\theta}$ is the ensemble average of the target Hamiltonian calculated over the samples generated with the distribution $P_{\theta}(\mathbf{x})$. An ansatz capable of modeling the probability distribution $P_{\theta}(\mathbf{x})$ is the Recurrent Neural Network (RNN) \cite{RNN1,RNN2}. Its elementary building block is the recurrent cell defined at each site (see FIG. \ref{fig:VNA_protocol}), which nonlinearly maps the concatenation of the input vector $x_{i-1}$ and previous hidden vector $h_{i-1}$ to an output hidden vector $h_i$ as
\begin{equation}\label{eq:RNN_nonlinear}
    h_i = f(W_i[h_{i-1};x_{i-1}]+b_i),
\end{equation}
where $f$ is the nonlinear mapping, $W_i$ and $b_i$ are the corresponding weight matrix and bias vector of the recurrent cell at site (or asset) $i$. The hidden vector $h_n$ defined between two consecutive sites $n$ and $n+1$ encodes all the information about the previous configurations $\{x_i\}_{i\leq n }$. This enables the RNN to efficiently model the probability distribution function of strongly correlated systems, such as the PO Hamiltonian. For in-depth details about the architecture of RNN pertaining to the simulation of correlated systems, we refer the reader to~\cite{RNN1,RNN2}. The dimension of the hidden vector signifies RNN's ability to simulate correlated systems and is aptly known as the number of memory units, $N_{\rm units}$. Note that $x_i$ in Eq.~(\ref{eq:RNN_nonlinear}) denotes the one-hot encoding of the portfolio weights. The RNN models the joint probability distribution over $\mathbf{x}$ as a product of conditional probabilities,

\begin{equation}\label{eq:11}
    P_{\theta}(\mathbf{x}) = \prod_{i=1}^N P_{\theta}(x_i|x_{i-1}.\ldots,x_1).
\end{equation}
FIG.~\ref{fig:VNA_protocol} shows details of the VNA algorithm. At $t =0$ we begin at temperature $T(0) \gg 0$ and perform a series of warm-up steps to ensure the system is well equilibrated at that temperature, i.e. $P_{\theta,t=0}(\mathbf{x}) \sim e^{-\frac{H(\mathbf{x})}{T(0)}}$. This is followed by a sequence of annealing steps with a predefined annealing schedule to bring the temperature to zero $T(t = 0) \rightarrow T(t = 1) = 0$. Throughout this paper, we use a geometric annealing schedule for the temperature. Each annealing step has several gradient descent steps to equilibrate the system at the present temperature. At $T=0$, the probability distribution $P_{\theta^*,t=1}(\mathbf{x})$ is expected to sample the ground state of the target Hamiltonian. Each gradient descent step involves a forward pass where a set of portfolio weights are sampled from the probability distribution $\{\mathbf{x}, P_{\theta}(\mathbf{x})\}$, these synthetic data are used to calculate the cost function $F_{\theta}(\mathbf{x})$ defined in Eq.~(\ref{eq:10}), and finally the gradients are calculated. The weights are updated during the backward pass $\theta(t+\delta t) = \theta(t) + \alpha \nabla_{\theta}F_{\theta,t}(\mathbf{x})$. Finally, at the end of the annealing, the RNN is trained and updated with the final set of parameters $\{\theta^*\}$. The network can now be used to generate uncorrelated samples $\{\mathbf{x},P_{\theta^*}(\mathbf{x})\}$ that optimize the target Hamiltonian. Not all samples generated with the distribution $P_{\theta^*}(\mathbf{x})$ will satisfy all financial constraints which is possible only in the limit of $\rho_n,\rho_{\nu},\rho_c,\rho_t \rightarrow \infty$. Therefore, the samples are passed through a filter and only those samples that satisfy all the constraints are chosen. For a more detailed explanation of the VNA method, we refer the interested reader to~\cite{VNA}.

\section{Results} \label{sec:resutls}
\subsection{Financial Data Preparation}\label{subsec:findataprep}

In this section, we present results on different financial indices. Our dataset is made up of historical prices over ten years of the S\&P 500, Russell 1000 and Russell 3000 indices. Only assets that are consistently present during the entire trading period are considered. Consequently, we have 478, 857 and 2008 respective stocks in the aforementioned indices. Closing prices with monthly granularity are used to compute expected returns and risk terms. 
$\bm{\mu}$ is computed as a logarithm returns, since it is time-additive, and the covariance matrix $\mathbf{C}$ is built with a five-year look-back window. The previous month's portfolios used to evaluate the transaction costs in Eqs.~(\ref{eq:4}) and (\ref{eq:5}) were obtained by solving the PO problem without transaction costs but with turnover, normalization, and volatility constraints using Mosek. The equally weighted portfolio is used in the turnover term in Eq.~(\ref{eq:9}).

\subsection{Scaling Analysis on the S\&P 500}\label{subsec:scaling_SnP500}

\begin{figure}
    \centering
    \includegraphics[width=\linewidth]{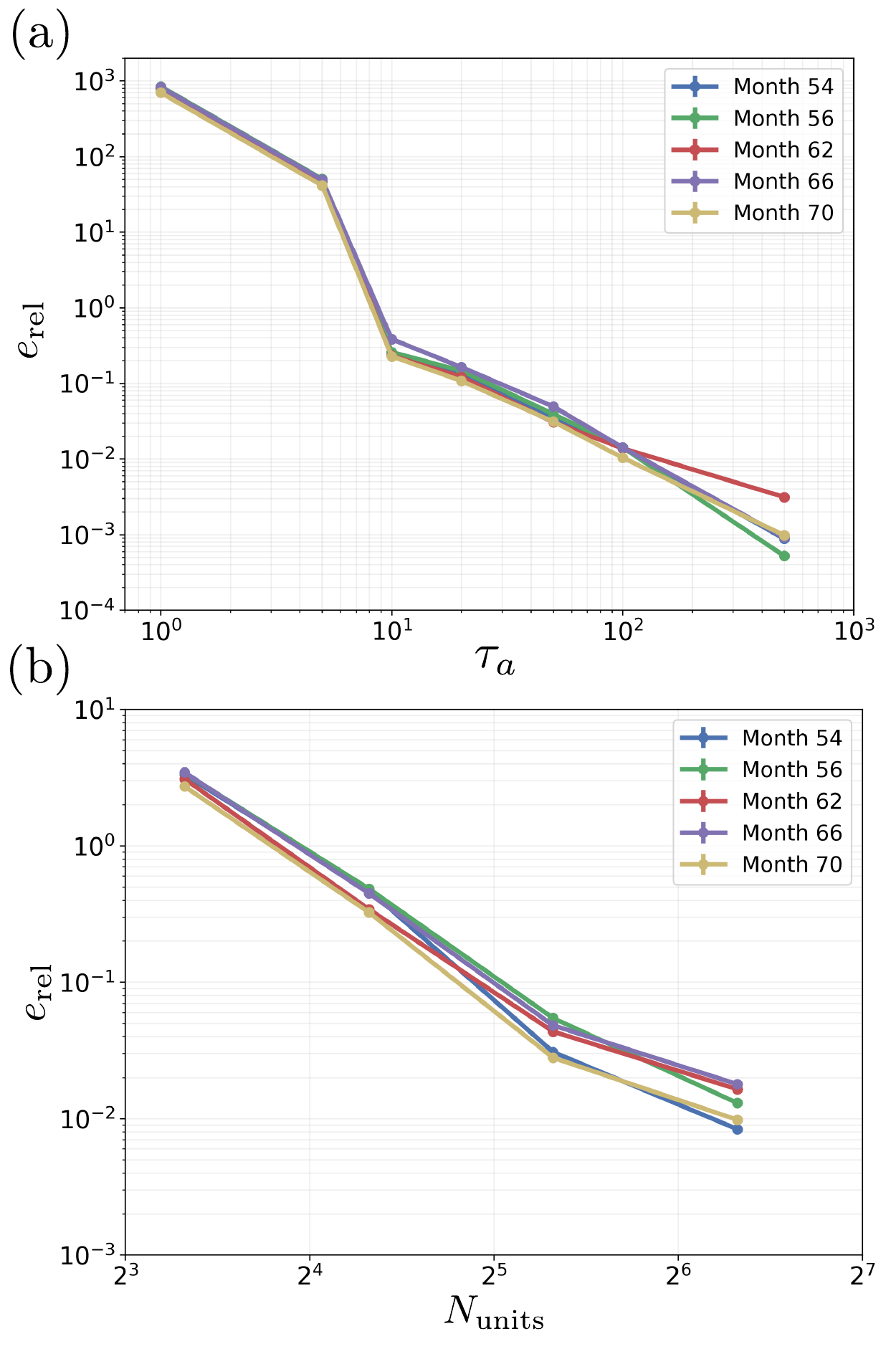}
    \caption{ (a) Relative energy $e_{\text{rel}}$ at the end of the VNA simulations as a function of annealing steps $\tau_a$. Simulations are performed at different trading months on the S\&P 500 index. The number of stocks in S\&P 500 is set to 478. The dimension of the RNN hidden state, i.e., the number of units $N_{\rm units}$ is fixed to 80. (b) $e_{\text{rel}}$ at the end of VNA simulations as a function of the number of units $N_{\rm units}$ of the RNN. The number of annealing steps is fixed to 50. Error bars are computed over the number of training samples of the RNN at the end of annealing.}
    \label{fig:VNA_energy_scaling}
\end{figure}

We start this section by presenting results of the VNA algorithm on the MINLP formulation of the portfolio optimization problem on the S\&P 500 as a function of annealing steps and parameters of the RNN ansatz. In FIG.~\ref{fig:VNA_energy_scaling}(a), we plot the relative energy,
\begin{equation}\label{eq:12}
    e_{\text{rel}} = \frac{|\langle H(\mathbf{x}) \rangle_{\theta} - E_{\rm ref}|}{|E_{\rm ref}|},
\end{equation}
where $\langle H(\mathbf{x}) \rangle_{\theta} $ denotes the objective function average obtained from samples generated at the end of the annealing procedure for a given number of $\tau_a$ annealing steps. $E_{\rm ref}$ is chosen as the lowest objective function value in a large number of annealing steps, which is set to $\tau_a=1000$. We use this definition of $e_{\text{rel}}$ as a proxy for the residual energy metric~\cite{king2023quantum,chowdhury2025}, which is often utilized to probe the ability to find low-energy solutions in disordered Ising models. 

In FIG.~\ref{fig:VNA_energy_scaling}(a), we present results for five different trading months to mimic the notion of different disorder realizations of couplings in spin glass theory. We note that, similar to disordered Ising models, $e_{\text{rel}}$ decreases with respect to $\tau_a$, thus indicating that the annealing paradigm is useful to find portfolios with lower objective values. Furthermore, we observe that, above $\tau_a=10$, VNA displays a power-law decay of the relative energy for different trading periods. This is reminiscent of what the Kibble-Zurek mechanism predicts, which was also observed in simulated (quantum) annealing, quantum annealing, and variational neural annealing of disordered spin glasses~\cite{king2023quantum, VNA}. Following the usual procedure of disordered averaging and taking averages across different trading months, we obtain $e_{\text{rel}} \propto {\tau_a}^{-1.43(1)}$ (see Appendix~\ref{sec:appendixa}). We note that this exponent should be interpreted rather qualitatively, as proper disordered averages should be taken over hundreds to thousands of different disorder realizations. However, this is prohibitive in our current setting due to the shared cost of a single VNA run.

FIG.~\ref{fig:VNA_energy_scaling}(b) instead displays the dependence of $e_{\text{rel}} $ with the number of hidden units $N_{\rm units}$ of the RNN cell. In this case, the number of annealing steps is set to 50, and $E_{\rm ref}$ is taken to be the lowest objective function value in the largest number of units considered for this setting. We observe that the model becomes more accurate as it is made to be more expressive. This is an interesting observation, as it seems to point to the fact that autoregressive models more powerful than RNNs will likely achieve a better accuracy in solving the PO problem within the VNA setting, as was recently shown for spin glass Hamiltonians~\cite{biazzo2024sparse, ma2024message}. Note here that we were unable to find a scaling dependence on $N_{\rm units}$ as was the case for $\tau_a$. It is very likely that increasing further $N_{\rm units}$ will lead to a plateau in relative energy, as the neural network would likely have reached enough capacity to represent the probability distribution of near-optimal solutions.

\begin{figure}
    \centering
    \includegraphics[width=\linewidth]{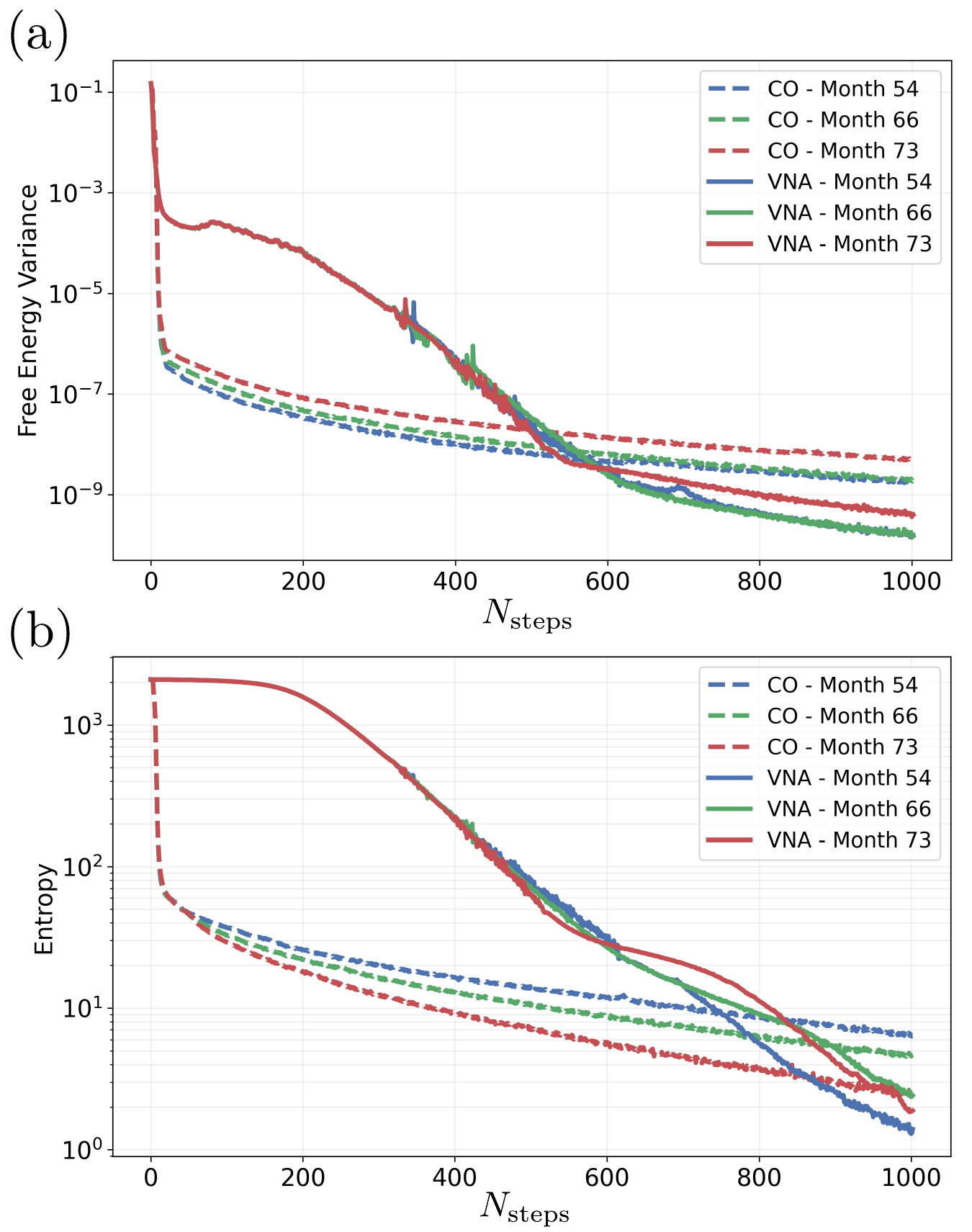}
    \caption{(a) Variance of the free energy and (b) Von-Neumann entropy as a function of $N_{\text{steps}}$ gradient descent steps at different trading months on the S\&P 500 index. The dotted lines represent Classical Optimization (CO) of the portfolio optimization free energy at zero temperature (i.e, without annealing). The solid lines represent VNA simulations. The number of stocks in S\&P 500 is set to 478.}
    \label{fig:VNA_CO_varE_entropy}
\end{figure}


\begin{figure}
    \centering
    \includegraphics[width=\linewidth]{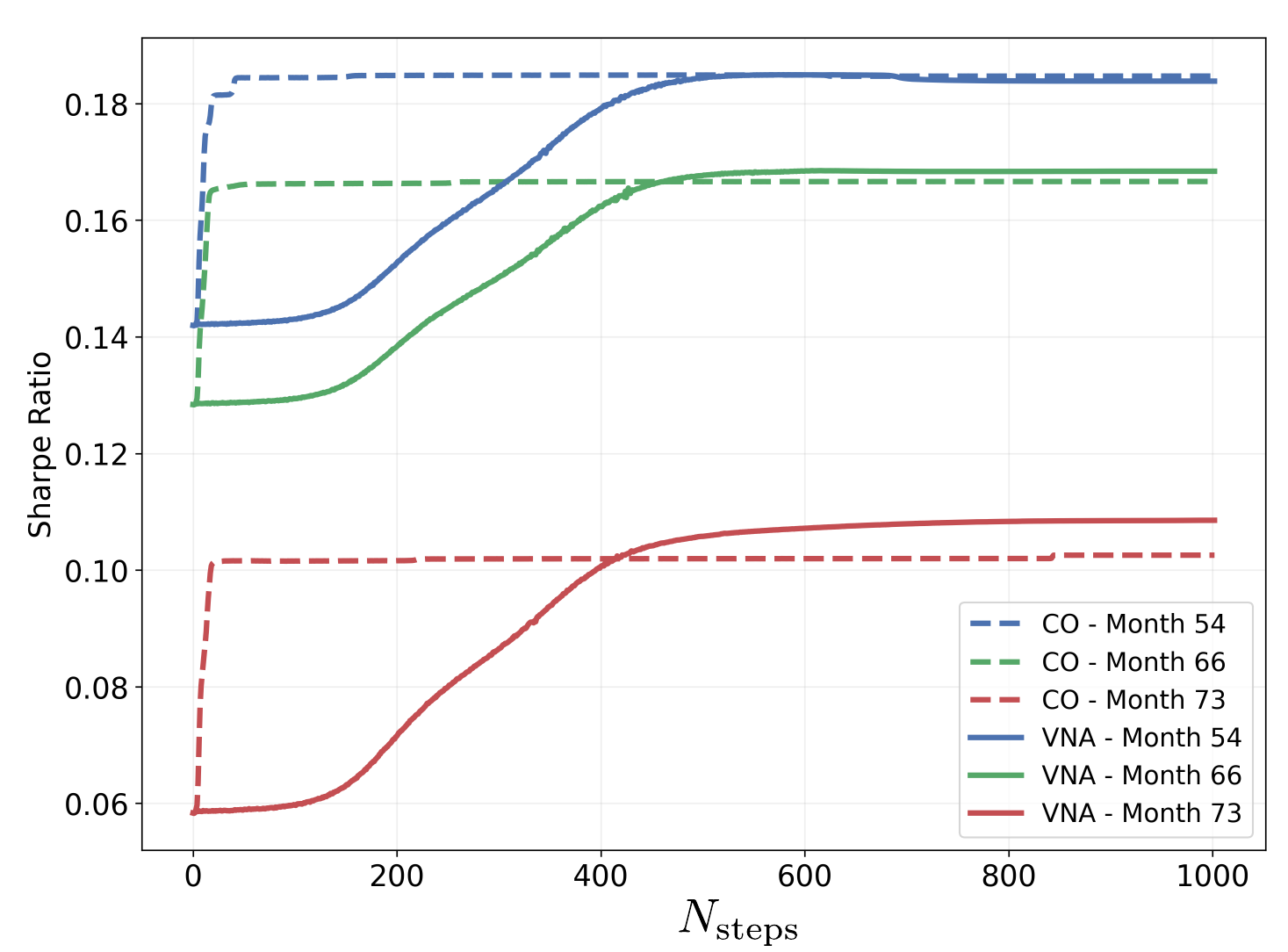}
    \caption{The average Sharpe ratio as a function of $N_{\text{steps}}$ gradient descent steps for CO and VNA simulations presented in FIG.~\ref{fig:VNA_CO_varE_entropy}. The Sharpe ratio is evaluated without a risk-free rate and transaction costs.}
    \label{fig:VNA_CO_sharpes}
\end{figure}

\subsection{Classical Optimization Versus VNA}
In this section, we perform an annealing ablation study to determine how important the annealing procedure is in finding low-energy solutions to the PO MINLP problem. We first turn off the annealing schedule, effectively performing a direct classical optimization (CO) of the objective function. We also solve the same PO problem with VNA. We compute two different metrics to determine the accuracy of the simulations: the variance of the variational free energy and the Von-Neumann entropy. Note that for the case of CO, although the entropy term is not explicitly used to solve the optimization problem, it serves as a good indicator of how the simulations proceed to find low-energy solutions. Both quantities are important to monitor, especially for the prospect of solving real-world optimization problems where the solution is not known a priori, as both metrics stand as good indicators of a successful optimization, irrespective of the objective function. 

FIG.~(\ref{fig:VNA_CO_varE_entropy})(a) shows the dependence of the variance of the free energy with respect to the number $N_{\rm steps}$ of gradient descent steps. For both CO and VNA, $N_{\rm steps}$ indicates the number of times the model parameters are updated, regardless of whether annealing is performed or not. As before, simulations are performed on the S\&P 500 index at different trading months. As expected, we observe that the variational free energy decreases with $N_{\rm steps}$ with VNA achieving lower values at a large number of steps. Here, the VNA is performed with $\tau_a=200$ and five equilibrium steps in between temperature updates. Another interesting observation is that CO finds low-energy solutions very quickly, as it reduces the variance by almost six orders of magnitude after only a few gradient descent steps. However, it gets stuck on those local minima for very long times, as indicated by the fact that later it takes almost three orders of magnitudes in $N_{\rm steps}$ to lower the free energy variance by two orders of magnitude. This slow convergence of direct variational optimization of classical Hamiltonians has also been observed in disordered Ising models~\cite{VNA} and protein folding~\cite{khandoker2025latticeproteinfoldingvariational}. 

FIG.~(\ref{fig:VNA_CO_varE_entropy})(b) further illustrates the two-stage behavior of CO. Similarly to the free energy variance, the Von-Neumann entropy displays a sharp initial drop followed by a very slow decrease in entropy, eventually achieving a higher level of entropy at the end of the optimization when compared with VNA which instead seems to have more exploratory-exploitative behavior compared to CO. Because the system is initially at high temperature, it has ample entropy. It can therefore perform more exploration of the free energy landscape. This is also shown in the relative high free energy variance it has at low $N_{\rm steps}$ compared to CO. However, in later stages, it can exploit the low-energy solution space more efficiently, eventually achieving lower variance and entropy compared to those of CO. These results highlight the importance of annealing in finding lower-energy solutions compared to direct optimization.

To investigate how the annealing advantage translates in a practical way to solve the problem of financial portfolio optimization, we monitor the Sharpe ratio defined in Eq.~(\ref{eq:3}) during the simulations performed in FIG.~\ref{fig:VNA_CO_varE_entropy}. FIG.~(\ref{fig:VNA_CO_sharpes}) shows that the Sharpe ratio is learned as the RNN is optimized in both CO and VNA.   We note that the rapid decrease in both free energy variance and the Von-Neumann entropy translates into a rapid increase in the Sharpe ratio for CO. However, the slow dynamics observed in both metrics result in the Sharpe ratio plateauing at the given value, with little to no improvement as the optimization is carried out.  VNA, on the contrary, has a more gradual increase in the Sharpe ratio with $N_{\rm steps}$, eventually outperforming the one obtained with CO. Note also that, though one may be tempted to use CO to obtain solutions to the MINLP quickly, we have observed that large number of gradient descent steps are still needed to find solutions that satisfy all the constraints.

\begin{figure}
    \centering
    \includegraphics[width=\linewidth]{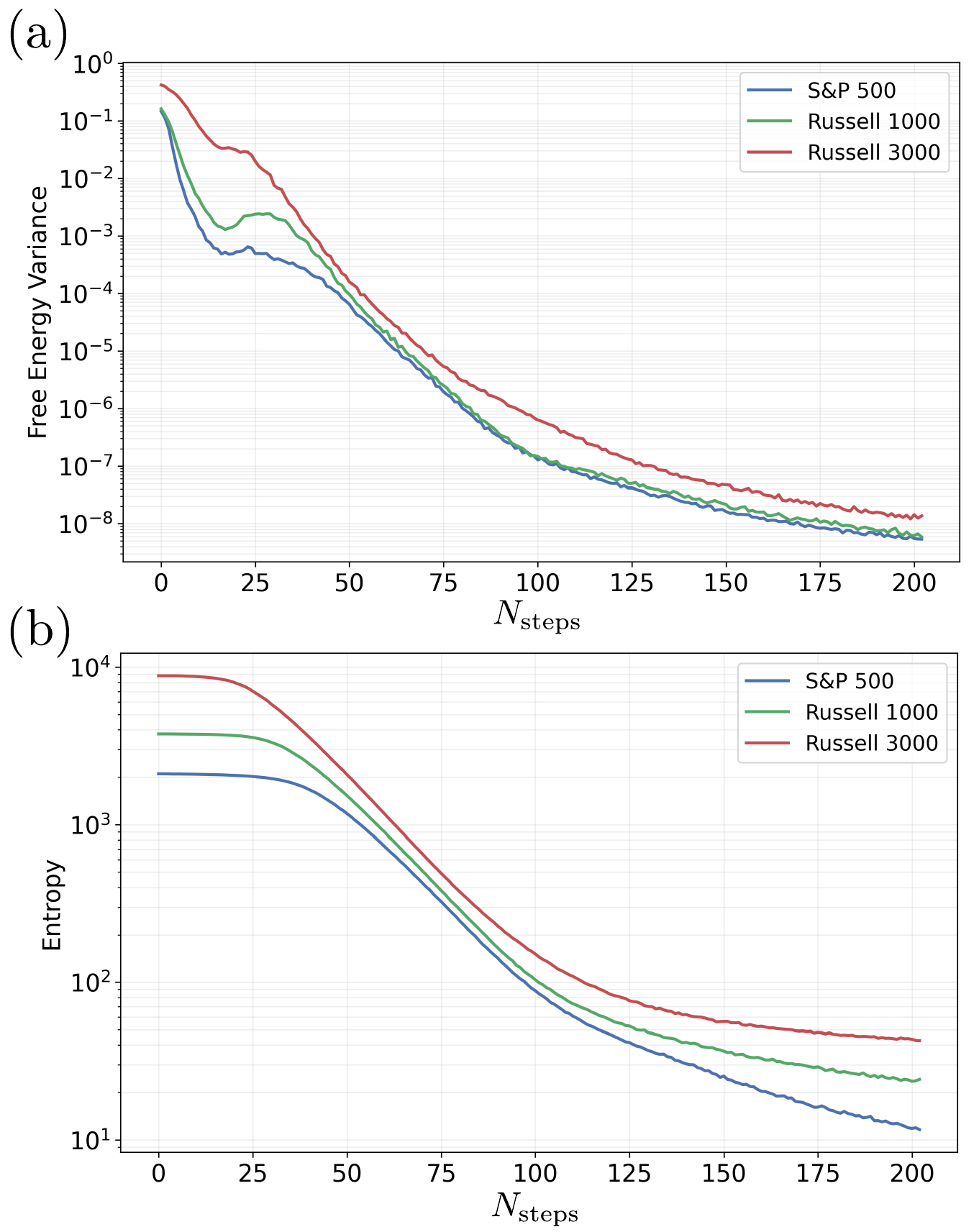}
    \caption{(a) Variance of the free energy and (b) Von-Neumann entropy as a function of $N_{\text{steps}}$ gradient descent steps for S\&P 500 (blue solid line), Russell 1000  (green solid line) and Russell 3000  (red solid line) indices for the same trading month. The number of stocks in S\&P 500, Russell 1000 and Russell 3000 is respectively 478, 853 and 2008.}
    \label{fig:VNA_Mosek_varF}
\end{figure}


\subsection{VNA Versus Mosek on Large-scale Portfolios}\label{subsec:mosekvsvna}
\begin{figure}
    \centering
    \includegraphics[width=\linewidth]{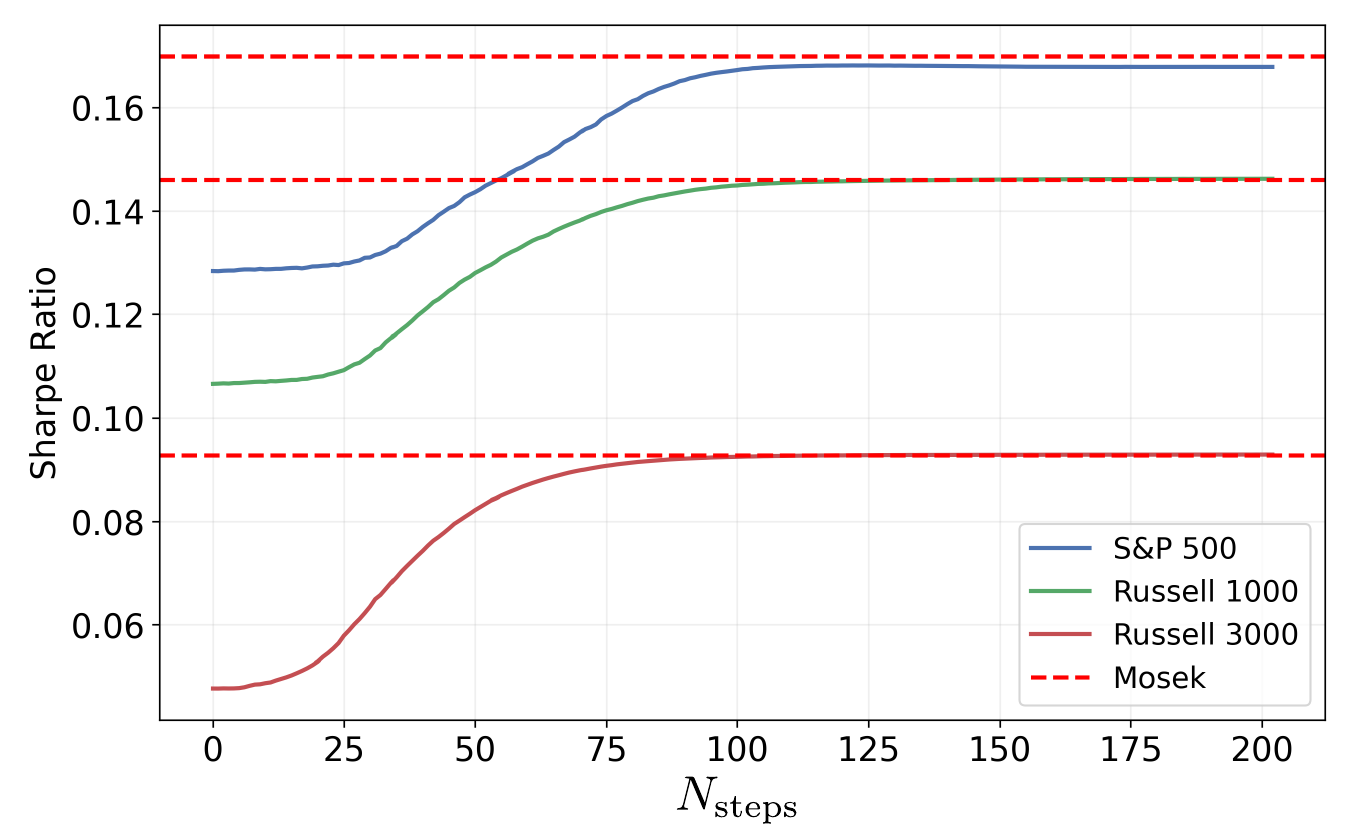}
    \caption{The average Sharpe ratio corresponding to simulations shown in FIG.~\ref{fig:VNA_Mosek_varF}. The red dotted lines are results from the mixed-integer solver Mosek obtained when solving for the corresponding objective functions for S\&P 500, Russell 1000 and Russell 3000, respectively. The Sharpe ratio is evaluated without a risk-free rate and transaction costs.  Comparisons on Time-to-Solution are shown in Table~\ref{tab:my_label}.}
    \label{fig:VNA_Mosek_sharpes}
\end{figure}

In this section, we study the behavior of the VNA method on large-scale portfolios and benchmark it against a commercial MINLP solver, specifically Mosek. As in the previous sections, we perform single-trading period optimization of financial portfolios, but now, in addition to the S\&P 500, we solve the PO on the Russell 1000 and Russell 3000 indices, with 857 and 2008 stocks, respectively. In FIG.~\ref{fig:VNA_Mosek_varF}, we observe that VNA still performs as before on Russell 1000 and Russell 3000 indices, as shown by the decrease in the free energy variance and entropy during annealing. This further demonstrates the ability of VNA to find low-energy solutions of the PO objective function even in the presence of non-linearity and non-convexity. Furthermore, FIG.~\ref{fig:VNA_Mosek_sharpes} shows that the Sharpe ratio gradually increases during annealing, and its value converges towards the one obtained by Mosek. This shows the ability of VNA to effectively learn financial metrics during the annealing process for large-scale portfolios. 

To compare the performance of VNA against Mosek, we analyze the time-to-solution to obtain the results presented in FIG.~\ref{fig:VNA_Mosek_sharpes}. Before discussing the results, we recall that the runtime of the VNA algorithm depends on several factors. Among them, the most critical is the number of annealing steps, assuming a fixed number of equilibration steps between subsequent temperatures. Other factors include the computational complexity of VNA, which is of $\order{N}$ for each gradient descent step when implemented with a RNN. The number of parameters in the RNN model, and the number of training samples are important factors too. 
In FIG.~\ref{fig:VNA_Mosek_varF}, these parameters were fixed to achieve results close to those of Mosek. Another important practical factor is the hardware on which the simulations are run, with more high-end Graphical Processing Units (GPUs) expected to produce shorter time-to-solutions. As such, the time-to-solution reported here should be viewed more qualitatively rather than quantitatively, which may differ from the definition typically used in similar studies. Nevertheless, this can still point to interesting results, particularly when considering instances of hard-to-solve problems. Regarding the evaluation of Mosek's time-to-solution, we consider the standard one used in MIP which is the time it takes to reach a relative gap of 0.01\%. For MIP solvers, this time can vary substantially depending not only on the difficulty of the objective function instance but also on the number of threads used in the CPU parallelization. The latter often has a non-linear dependence on the number of threads as shown in FIG.~\ref{fig:vna_vs_mosek_time_sharp} of Appendix~\ref{sec:appendixb}. Note also that, unlike deep learning models, MIP solvers such as Mosek, CPLEX~\cite{cplex}, or Gurobi~\cite{gurobi} currently do not support GPU acceleration for solving MINLP.
In Table~\ref{tab:my_label}, we show the time-to-solution of a hard-to-solve instances whose results were presented in Figs.~\ref{fig:VNA_Mosek_varF} and ~\ref{fig:VNA_Mosek_sharpes}. Notably, we highlight the case of Russell 3000, for which the actual stopping criterion was changed from 0.01\% to 1\% (hence the asterisk in Table~\ref{tab:my_label}) because the former threshold was not satisfied even after several days of simulations. Moreover, increasing the relative gap requirement by two orders of magnitude still resulted in Mosek having twice the time-to-solution as VNA. These results highlight not only the ability of VNA to provide faster near-optimal solutions for MINLP compared to commercial solvers on hard-to-solve instances but also demonstrate the limitations of such solvers in tackling real-world problems in an actual business setting, as the complexity of a given instance is not known beforehand. 

\begin{figure}
    \centering
    \includegraphics[width=\linewidth]{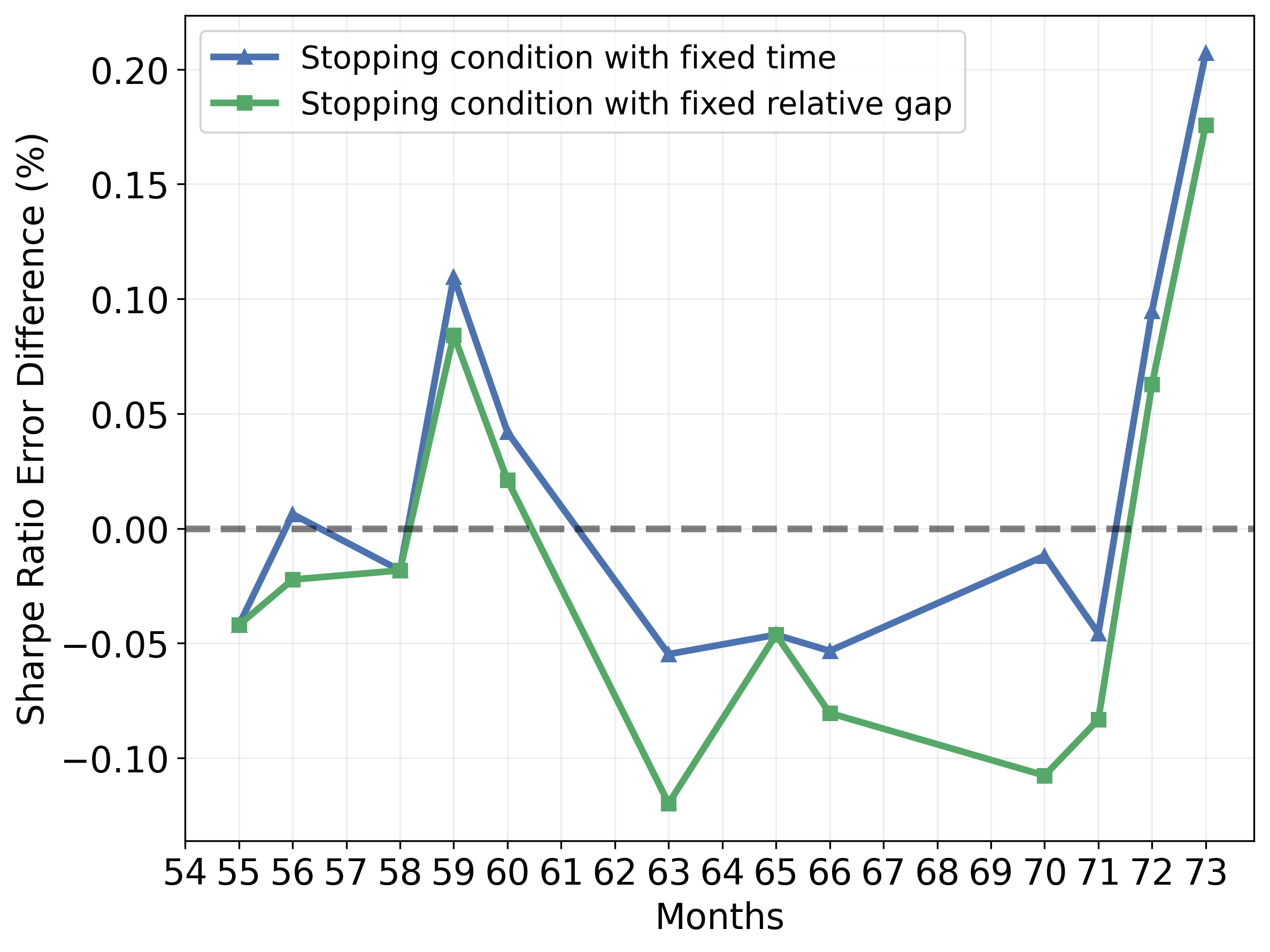}
    \caption{Percentage difference between the Mosek Sharpe ratio and the best valid VNA Sharpe ratio at the end of the VNA simulations on the Russell 3000 across different trading months. The blue triangle data points are obtained by setting Mosek’s stopping condition equal to the runtime of VNA in Table~\ref{tab:my_label}. The green square data points are obtained by setting Mosek’s stopping criterion to a relative gap of 5\%. Sharpe ratios are computed without a risk-free rate but include transaction costs. The black dotted line serves as a visual reference; positive values above it indicate Mosek superiority over VNA, while negative values indicate the reverse.  }
    \label{fig:VNA_Mosek_relative_sharpes}
\end{figure}

In addition to the time-to-solution, we also compare the quality of solutions produced by VNA and Mosek. For VNA, after generating an ensemble of portfolios via autoregressive sampling, we select the one with the highest Sharpe ratio that also satisfies all the constraints. To obtain a more realistic evaluation of the Sharpe ratio, we deduced both fixed and variable transaction costs from the expected returns in Eq.~\ref{eq:3}. The result is compared with the one obtained by Mosek. In FIG.~\ref{fig:VNA_Mosek_relative_sharpes}, we plot the difference in Sharpe ratios between Mosek and VNA (in percentages) across different trading months on the Russell 3000. Positive values indicate Mosek's superiority over VNA, while negative values indicate the opposite. Two stopping criteria are used for Mosek: the first (blue curve), is obtained by setting Mosek’s stopping condition equal to the runtime of VNA in Table~\ref{tab:my_label}, and the other (green curve), is obtained by setting Mosek’s stopping criterion to a relative gap of 5\%. The first stopping condition seeks to benchmark the solution quality when VNA and Mosek have similar runtimes, whereas the second condition seeks to set up Mosek in a setting more realistic to routine asset management operations where time-to-solution is very important. For all trading months, the second stopping condition yielded results within 18 seconds. In both cases, the Sharpe ratio difference is minimal, indicating that VNA can yield comparable results to Mosek across different trading periods, sometimes with slightly better results. The results obtained with a fixed relative gap of 5\% lead to a general reduction of the Sharpe ratio error difference. This is somehow expected as imposing such a high relative gap leads to fast solutions with lower Sharpe ratios whereas longer simulations lead to lower relative gaps hence better Sharpe ratios.
We note also that VNA was not able to find valid solutions for all the trading months in FIG.~\ref{fig:VNA_Mosek_relative_sharpes}. This is likely because the penalty condition fell short to satisfy all the constraints simultaneously (see Appendix~\ref{subsec:valid_solutions}).

\begin{table}
    \centering
\caption{Time to Solution (in seconds) of FIG.~\ref{fig:VNA_Mosek_sharpes}}
\label{tab:my_label}
    \begin{tabular}{|c|c|c|c|} \hline 
         &  \textbf{S$\&$P 500}&  \textbf{Russell 1000}& \textbf{Russell 3000}\\ \hline 
        \textbf{ VNA}&  625&  1331& 2580\\ \hline 
        \textbf{ Mosek}&  913&  6850& 6709$^{*}$\\ \hline
    \end{tabular}
\end{table}

\subsection{Critical Portfolio Dynamics}\label{subsec:critical}

In Section~\ref{subsec:scaling_SnP500}, we observed a power-law decay of the relative energy $e_{\text{rel}}$ as a function of the number of annealing steps. This scaling relation highlights the efficacy of VNA in progressively targeting lower energy solutions with an increasing number of annealing steps. We therefore ask whether this behavior follows a universal finite-size scaling across portfolios of different sizes. Several previous studies have successfully implemented the dynamical finite-size scaling (DFSS) ansatz~\cite{PhysRevB.84.224303} to study universal critical dynamics and scaling advantage in different paradigmatic hard optimization problems, including the spin-glass problem~\cite{PhysRevLett.134.160601,chowdhury2025,king2023quantum}. Adopting this framework, we treat $e_{\text{rel}}$ as a proxy of residual energy, mimic disorder averages in spin glass by averaging over different trading months, and treat the indices, S$\&$P 500, Russell 1000, and Russell 3000 with 478, 857, and 2008 stocks (see Section~\ref{subsec:findataprep} for details on data preparation) as three increasing system sizes. For this specific section, we have relaxed the constraint space only to include the normalization constraint. Specifically, in addition to the penalty constraint, we have also included a Lagrange multiplier term to enforce the normalization constraint (see Appendix~\ref{subsec:valid_solutions}). The objective Hamiltonian then takes the following form,
\begin{equation}\label{eq:criticality}
    H(\mathbf{x}) = -\bm{\mu}^T \mathbf{x} + \rho_{\text{v}}\, \mathbf{x}^T\mathbf{C} \mathbf{x} + \lambda_{\rm n} |\mathbf{1}^T \mathbf{x} - 1| +\rho_{\text{n}} \left(\mathbf{1}^T \mathbf{x} - 1\right)^2,
\end{equation}
where, $\lambda_{\rm n}$ is the Lagrange multiplier corresponding to the normalization constraint. The presence of the Lagrange multiplier term on top of the penalty term enforces feasibility and faster convergence ~\cite{bertsekas1997nonlinear,nocedal2006numerical}. 

To test the dynamical finite-size scaling, we introduce the DFSS ansatz,

\begin{equation}
    e_{\text{rel}} = N^{-\kappa} g(\tau_a N^{z+1/\nu}),
\end{equation}

where, $\kappa$, $z$, and $\nu$ are the critical exponents. Within this formulation if we rescale the relative energy $e_{\text{rel}}(N,\tau_a) \rightarrow e_{\text{rel}}(N,\tau_a)N^{k}$ and the annealing time $\tau_a \rightarrow \tau_a N^{z+1/\nu}$, all the data corresponding to different system sizes should collapse onto a common universal curve~\cite{PhysRevB.84.224303, PhysRevB.89.054307}. Consequently, the annealing time $\tau_a$ necessary for the system to remain adiabatic~\cite{VNA} scales as

\begin{equation}
    \tau_a \sim N^{-\mu}, \hspace{0.1cm} \text{where} \hspace{0.3cm} \mu = z+1/\nu.
\end{equation}

\begin{figure}
    \centering
    \includegraphics[width=\linewidth]{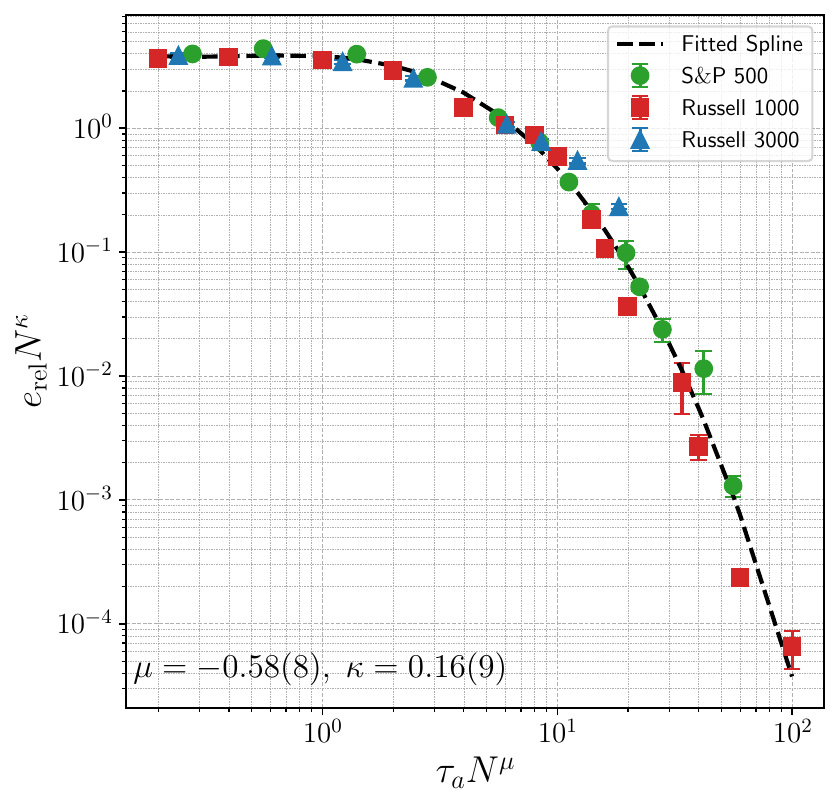}
    \caption{Finite size scaling collapse of the relative energy $e_{\text{rel}}$ as a function of annealing time $\tau_a$ for three different indices: S\&P 500, Russell 1000, and Russell 3000 (actual number of stocks 478, 857, and 2008, respectively). Each data point is averaged over five independent trading months (see FIG. \ref{fig:VNA_energy_scaling}). The black dashed line is the cubic univariate spline interpolation $g(x)$ obtained from the global fit. The concatenation of three data sets shows a single scaling function $e_{\text{rel}}N^k = g(\tau_a N^{\mu})$.}
    \label{fig:collapse}
\end{figure}

In FIG.~\ref{fig:collapse} we present the finite size scaling collapse of data corresponding to three indices with an increasing number of stocks into a standard universal curve. The critical exponents extracted from this collapse are $\mu = -0.58 \pm 0.08$ and $\kappa = 0.16 \pm 0.09$. The full details of the collapse technique and the errors can be found in the Appendix~\ref{subsec:collapse_error}. The relatively large uncertainties in critical exponents can be attributed to mainly two reasons: a limited number of data points, and each data point being a disorder average over only five trading months significantly fewer than hundreds to thousands of disorder average employed in spin-glass studies~\cite{chowdhury2025,king2023quantum,PhysRevLett.134.160601,VNA}. Beyond unraveling the underlying universal dynamics and thus bridging statistical physics with finance, the finite size collapse observed in FIG.~\ref{fig:collapse} also positions VNA as a robust and scalable method capable of forecasting the performance of larger future indices as a function of computational resources without the need to run expensive benchmarks.

\section{Conclusions} \label{sec:conclusions}

We have presented results of the variational neural annealing algorithm on large-scale portfolio optimization problems using real-world data from the S\&P 500, Russell 1000 and Russell 3000 indices. We have implemented the formulations of portfolio optimization problem with realistic terms and constraints, which resulted in a mixed-integer nonlinear program. We have demonstrated that, similar to previous results on disordered Ising glasses and combinatorial optimization problems, the VNA method, implemented with autoregressive models, can find lower-energy solutions to the portfolio optimization problem with increasing annealing time. In particular, results on different instances of the PO problem on the S\&P 500 reveal a power-law dependence of the residual energy with respect to the annealing time reminiscent of what is observed in spin glasses under the Kibble-Zurek mechanism. Through a systematic ablation approach, we have shown that the annealing paradigm is important for achieving lower objective values in financial portfolios compared to classical optimization. A behavior that was previously observed on disordered Ising models~\cite{VNA} and protein folding~\cite{khandoker2025latticeproteinfoldingvariational}. 

Additionally, we have also demonstrated that VNA can identify optimal solutions to large-scale equity portfolios comprising more than 2000 assets, even in the presence of mixed-integer nonlinear constraints. Physics-inspired metrics, such as the free energy variance and the Von-Neumann entropy, have been shown to be effective metrics for assessing the performance of VNA simulations. This is a preliminary step to provide a level of interpretability of the results and could be used as a regulatory metric, especially with the prospect of using VNA to solve real-world problems. Furthermore, VNA was shown to be competitive with a commercial solver, in this case Mosek, in terms of solution quality, while leading to a faster solution for hard-to-solve instances. We also performed a dynamical finite-size scaling of VNA on the S\&P 500, Russell 1000 and Russell 3000 indices and observed that VNA exhibited a universal scaling behavior, which suggests that the annealing time required for the system to remain adiabatic scales polynomially with the number of assets in the portfolio highlighting the scalability of VNA.

Several directions can be taken to improve the results presented in this article. One approach involves utilizing more powerful neural networks, such as those that employ a message passing algorithm~\cite{ma2024message} or those that directly encode the objective function parameters into the neural network~\cite{biazzo2024sparse}. Another direction would be to use natural gradient for training, as it was recently shown to outperform the Adam optimizer when using variational annealing to find solutions of spin glass Hamiltonians~\cite{liu2025efficient}. 

Finally, in Section~\ref{subsec:mosekvsvna}, we noted that VNA did not yield valid solutions for all trading months, primarily due to the limitations of standard penalty methods, which only guarantee constraint satisfaction in the limit of infinitely large penalty coefficients. While this poses challenges for optimizing the financial portfolio objective function, it also opens up promising avenues for improvement. While a remedy was shown to yield valid solutions in Section~\ref{subsec:critical} by adding a Lagrangian term to the objective function, it is not optimal for the prospect of solving different constraints as this will require expensive hyperparameter tuning. Therefore, a possible next step is to incorporate an augmented Lagrangian approach, which supplements the penalty terms with adaptive Lagrange multiplier updates~\cite{bertsekas1997nonlinear,nocedal2006numerical}. This strategy has been shown to promote both feasibility and convergence at moderate Lagrange multipliers and penalty coefficient values~\cite{kotary2024learning,basir2023adaptive}, leading to a favorable balance between objective optimization and constraint enforcement.


\section*{Acknowledgments}
We thank Mahmoud El Mabrouk, Hanna Morilhas, Roger Melko and Juan Carrasquilla for fruitful discussions. EMI acknowledges support from the Natural Sciences and Engineering Research Council of Canada (NSERC). yiyaniQ acknowledges support from the Perimeter Institute for Theoretical Physics. Research at Perimeter Institute is supported in part by the Government of Canada through the Department of Innovation, Science and Economic Development Canada and by the Province of Ontario through the Ministry of Economic Development, Job Creation and Trade. NR acknowledges support from the Mitacs Accelerate Umbrella program. Computer simulations were made possible thanks to the Digital Research Alliance of Canada cluster.
\appendix 
\section{Additional scaling analysis results on S\&P 500} \label{sec:appendixa}

\begin{figure}
    \centering
    \includegraphics[width=\linewidth]{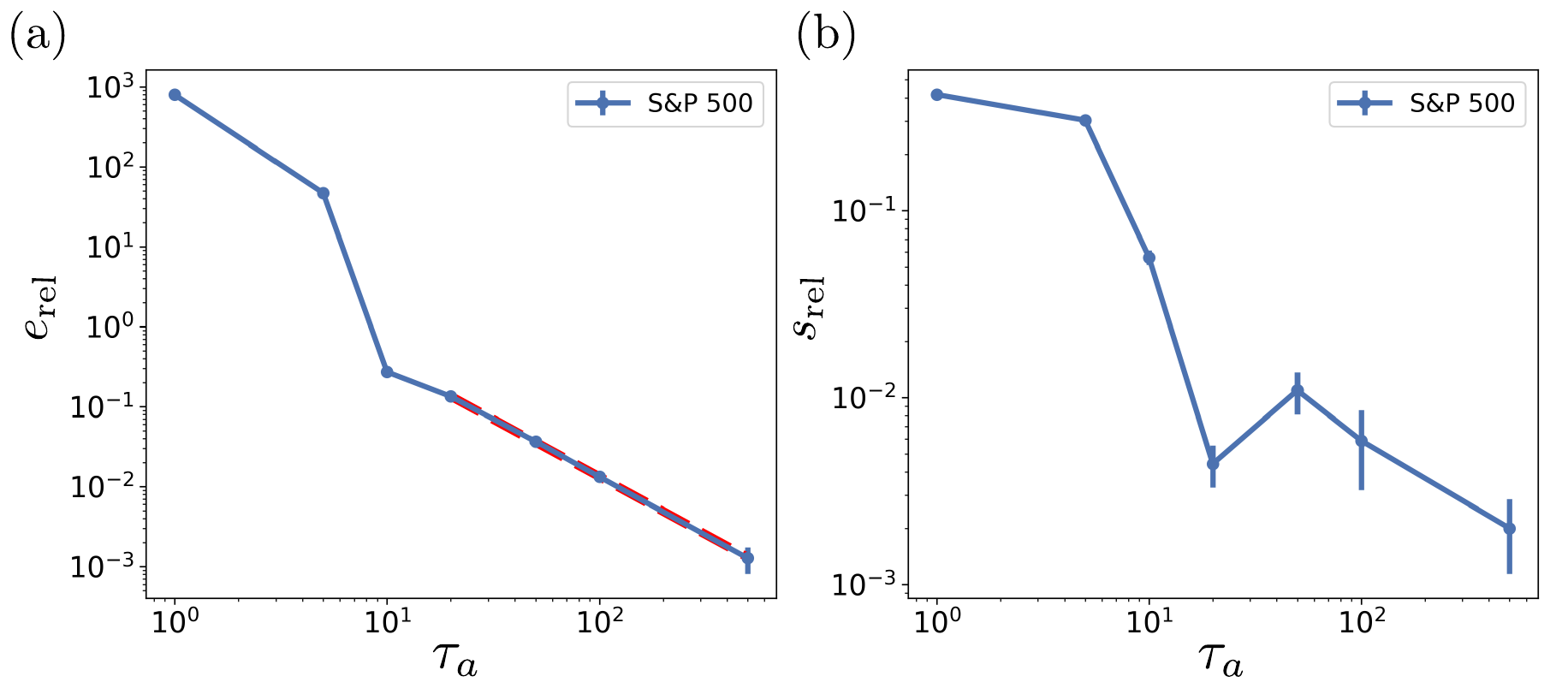}
   \caption{(a) Relative energy $e_{\text{rel}}$ at the end of the VNA simulations as a function of annealing steps $\tau_{a}$, averaged over the different trading months depicted in FIG.~\ref{fig:VNA_energy_scaling}(a).  The red dotted line represents a power-law fit over the last number of annealing steps. The relative error scaling is $e_{\rm rel} \propto {\tau_{a}}^{-1.43(1)}$. (b) Corresponding relative Sharpe ratios $s_{\text{rel}}$ averaged over the same trading months. }
    \label{fig:vna_energy_sharpe_Annealing_time}
\end{figure}

In this section, we present additional results on the scaling analysis of the VNA method on the S\&P 500 discussed in Section~\ref{subsec:scaling_SnP500}. In FIG.~\ref{fig:vna_energy_sharpe_Annealing_time}(a), we plot the relative energy scaling $e_{\text{rel}}$ averaged over the trading months shown in FIG.~\ref{fig:VNA_energy_scaling}(a). The trading months' average is used in analogy to disorder averages traditionally used in the context of Ising spin glasses. Despite the small number of trading months, five in this plot, two regimes are clearly identifiable in FIG.~\ref{fig:vna_energy_sharpe_Annealing_time}(a). The first one is when the number of annealing steps $\tau_{a} < 10$, which corresponds to a quench-type regime where the system is rapidly cooled from a high temperature to zero temperature. Above that, the annealing dynamics follows an adiabatic regime which is marked by the power-law dependence of the relative energy with respect to the number of annealing steps, $e_{\rm rel} \propto {\tau_{a}}^{-1.43(1)}$. In FIG.~\ref{fig:vna_energy_sharpe_Annealing_time}(b), we plot $s_{\rm rel}$, which is the corresponding Sharpe ratio relative error. It is computed by replacing the objective value with the Sharpe ratio in Eq.~(\ref{eq:12}). Averages are also taken over the same trading months as before. Here, too, there appears to be a depiction of both quench-like and adiabatic-like behaviors; however, the limited number of points does not permit the extraction of a scaling law. It is also interesting to note that the Sharpe ratio can be systematically improved with the number of annealing steps, albeit at the cost of higher computational times.

\begin{figure}
    \centering
    \includegraphics[width=\linewidth]{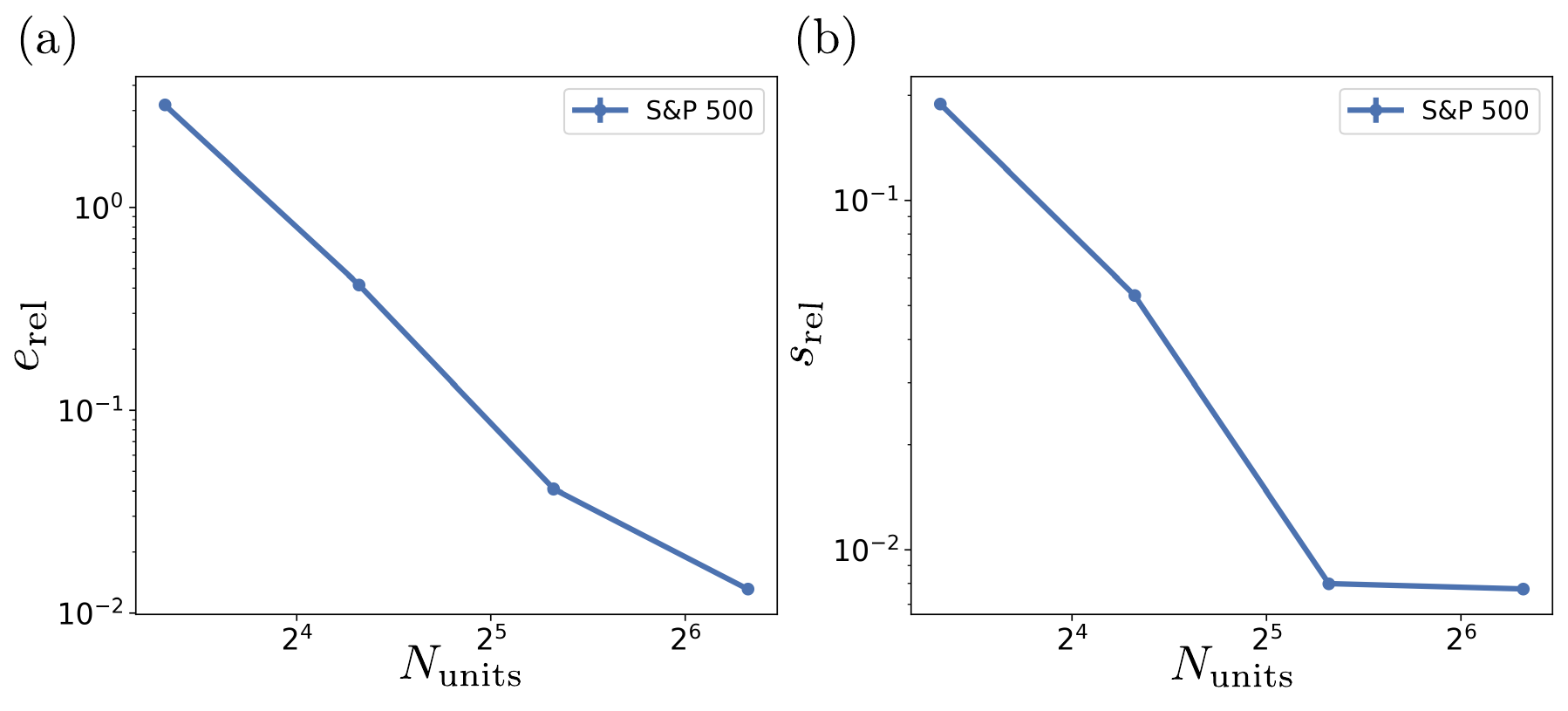}
    \caption{ (a) $e_{\text{rel}}$ at the end of VNA as a function of number of hidden units $N_{\rm units}$ of the RNN averaged over different trading months shown in FIG.~\ref{fig:VNA_energy_scaling}(b). (b) Corresponding $s_{\text{rel}}$ averaged over the same trading months. }
    \label{fig:vna_energy_sharpe_Nunits}
\end{figure}

The relative energy scaling $e_{\text{rel}}$ averaged over the trading months shown in FIG.~\ref{fig:VNA_energy_scaling}(b) is depicted in FIG.~\ref{fig:vna_energy_sharpe_Nunits}(a). Here, the plot also displays a systematic improvement of the energy relative error with respect to the RNN hidden state dimension $N_{\rm units}$. We note, however, that no scaling relation could be extracted, likely due to the limited number of available data points. FIG.~\ref{fig:vna_energy_sharpe_Nunits}(b) depicts the corresponding Sharpe ratio relative error $s_{\rm rel}$ averaged over the same trading months. Interestingly, we observe a plateau which was not observed for $e_{\text{rel}}$. This is because at higher values of $N_{\rm units}$, terms in the portfolio optimization objective function that do not appear in the Sharpe ratio are further optimized leading to a steady decrease in $e_{\text{rel}}$. We expect though that when $N_{\rm units}$ reaches a value where the RNN has enough representational power to model the low-energy solutions, $e_{\text{rel}}$ will also converge. 

\section{VNA Versus Mosek on S\&P 500 }
\label{sec:appendixb}
\begin{figure}
    \centering
    \includegraphics[width=\linewidth]{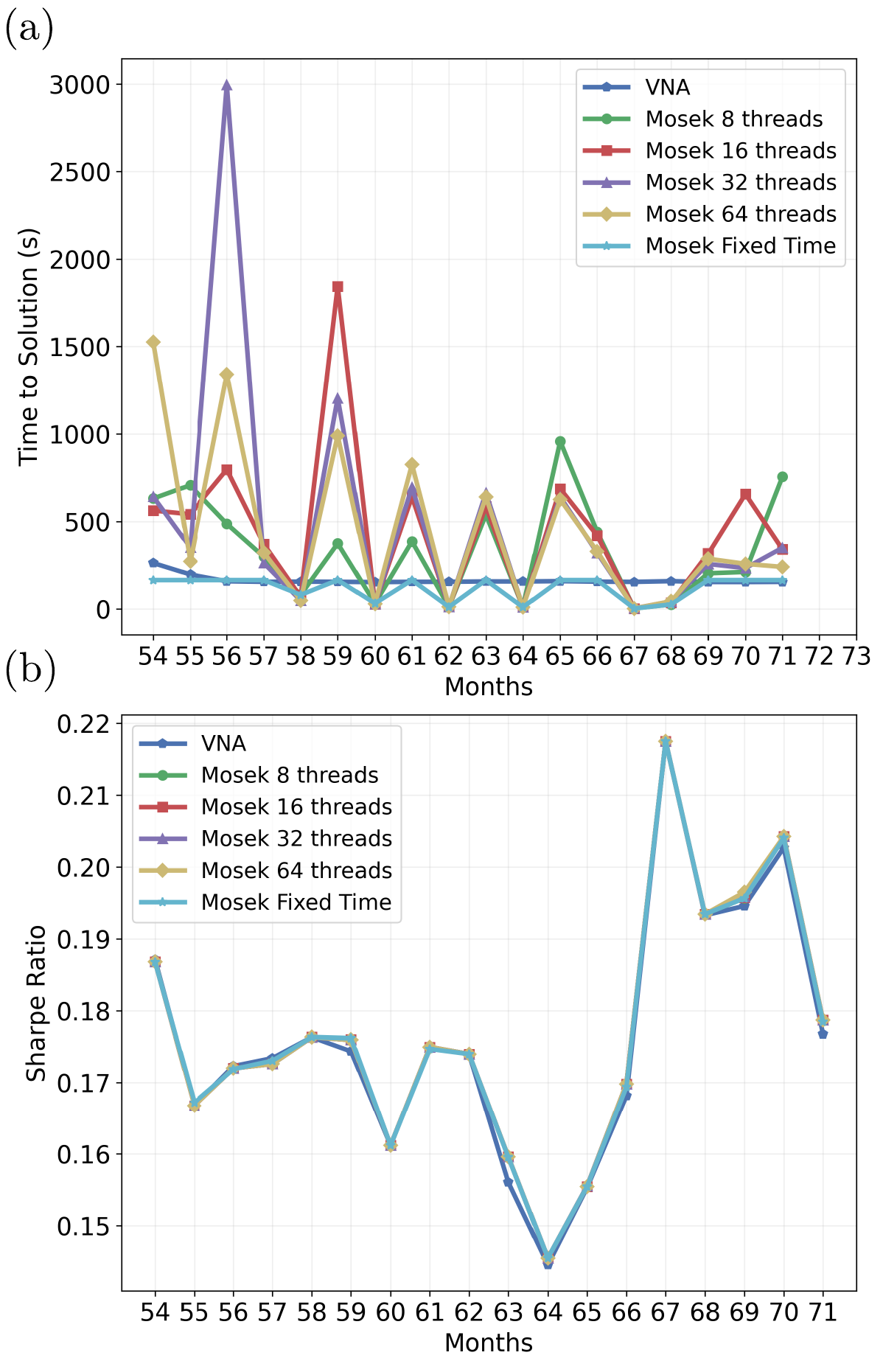}
    \caption{ (a) Time-to-solution (in seconds) for VNA and Mosek across various trading months on the S\&P 500. Mosek results are shown for different numbers of CPU threads at a fixed relative gap tolerance of 0.01\%. The light blue curve represents Mosek runs without a relative gap constraint, but limited to the maximum runtime of the corresponding VNA simulations. The number of annealing steps for VNA is fixed at 50. (b) Sharpe ratios corresponding to the solutions obtained in panel (a).
}
    \label{fig:vna_vs_mosek_time_sharp}
\end{figure}

In this section, we benchmark VNA against Mosek on the S\&P 500 for the MINLP problem discussed in Section~\ref{subsec:scaling_SnP500}. In FIG.~\ref{fig:vna_vs_mosek_time_sharp}(a), we plot the time-to-solution (as defined in the main text) of both methods for different trading months. For VNA, the simulations are run for $\tau_{a}=50$ steps. Mosek blue triangle data points use the corresponding fixed time as the stopping condition, as specified by VNA, and a standard relative gap error of 0.01\%. The other Mosek's data points have only the latter condition as a stopping criterion. Mosek's simulations are performed in parallel with multiple CPU threads to assess Mosek's dependence on the number of threads. FIG.~\ref{fig:vna_vs_mosek_time_sharp}(a) shows that a stopping condition with the standard relative gap error leads to different time-to-solution depending on the instance of the PO objective function used for a given trading period.

Furthermore, it also displays the non-linear dependence on the number of threads of the Mosek solver (e.g., Month 56). FIG.~\ref{fig:vna_vs_mosek_time_sharp}(b) shows that the Sharpe ratios of the simulations performed in FIG.~\ref{fig:vna_vs_mosek_time_sharp}(a) result in comparative results for VNA and Mosek (considering all stopping criteria). In particular, the Mosek's time criterion seems to be particularly appealing, but is not expected to remain a good criterion as the complexity of the optimization problem increases (e.g. trading whole shares instead of fractional ones) for which a valid solution may not be guaranteed to be found in a short amount of time. 

Unless otherwise stated, all Mosek simulations were performed on AMD processors with 64 CPU cores. VNA simulations were conducted using multiple GPUs (specifically, 4 NVIDIA A100 GPUs in parallel) except for Section~\ref{subsec:critical}, which utilized a single NVIDIA H200 GPU.


\section{Feasibility of solutions in VNA}\label{subsec:valid_solutions}
\begin{figure}
    \centering
    \includegraphics[width=\linewidth]{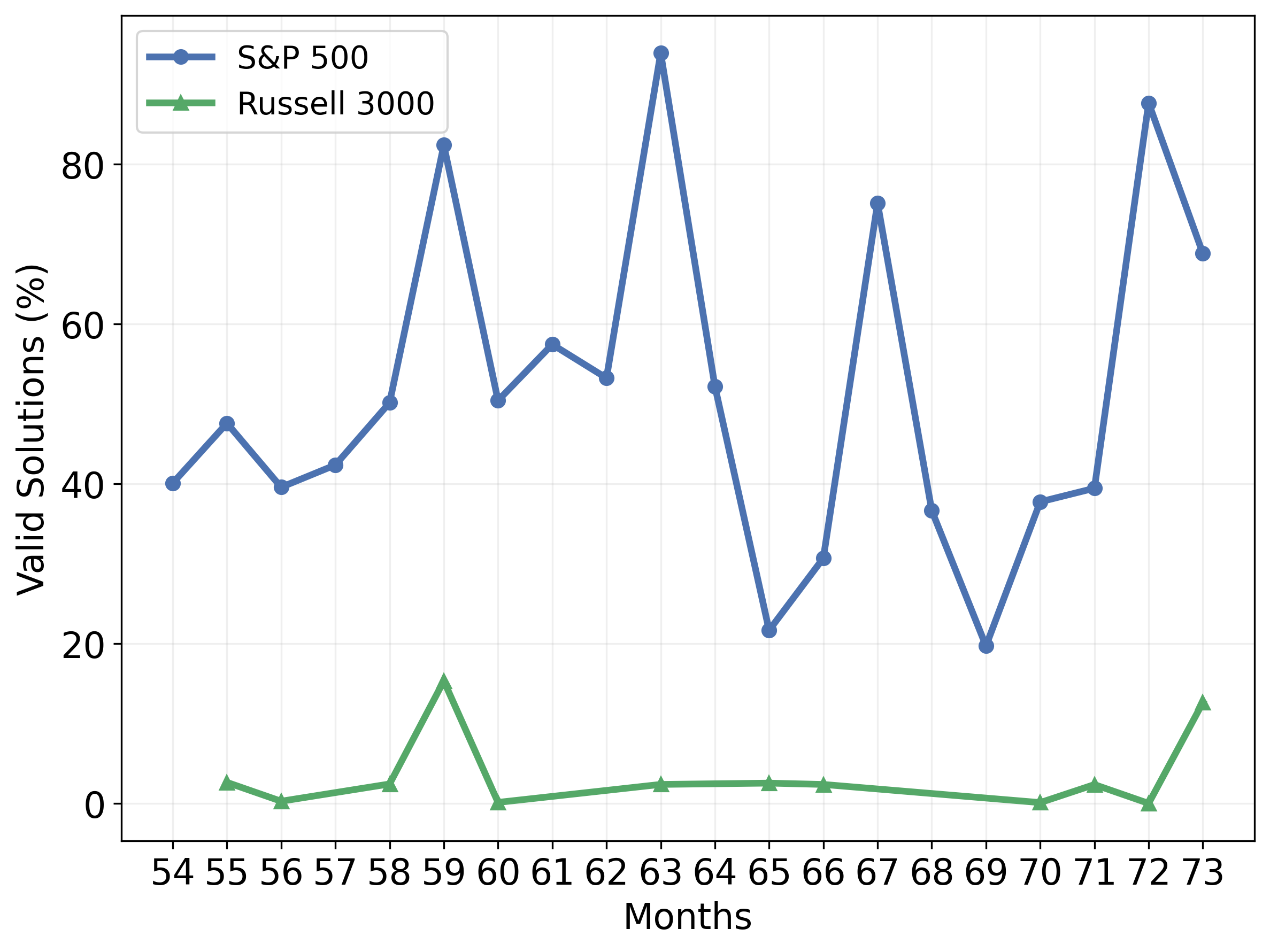}
    \caption{Percentage of valid solutions of VNA simulations with respect to different trading months. The blue curve depicts feasible solutions corresponding to the S\&P 500 data obtained in FIG.~\ref{fig:vna_vs_mosek_time_sharp} with $\tau_{a}=50$ steps. The green curve data corresponds to the ones of Russell 3000 presented in FIG.~\ref{fig:VNA_Mosek_relative_sharpes} where the number of annealing steps was set to $\tau_{a}=200$. Months without data points did not provide solutions that satisfy all the constraints.}
    \label{fig:vna_valid_solutions}
\end{figure}

This section assesses the capability of the VNA algorithm to generate feasible solutions to the portfolio optimization objective function in the presence of constraints. As a metric for feasibility, we compute the ratio of samples generated by the RNN that satisfy all constraints to the total number of samples generated. This ratio, expressed as a percentage, is evaluated at the end of the annealing process. As discussed in the main text, the feasibility of solutions strongly depends on the method used to enforce the constraints. In FIG.~\ref{fig:vna_valid_solutions}, constraint satisfaction is implemented via penalty terms applied to the normalization, turnover, and volatility constraints (see Section~\ref{sec:po}). 

The blue curve shows the percentage of valid solutions obtained for the S\&P 500 across various trading months, corresponding to the data in FIG.~\ref{fig:vna_vs_mosek_time_sharp}. VNA achieves up to 80\% validity across all months in this case. In contrast, the green curve shows results for the Russell 3000 over the same trading period, where the fraction of valid solutions deteriorates significantly, reaching as low as 0.129\%. Although the absolute number of valid solutions remains non-negligible due to the large sample size (36,000), the method does not consistently yield valid portfolios across all trading months, as evidenced by the presence of missing data points for some months. While it is theoretically possible to improve constraint satisfaction by tuning the penalty coefficients for the Russell 3000, this tuning itself constitutes a complex optimization problem.

\begin{figure}
    \centering
    \includegraphics[width=\linewidth]{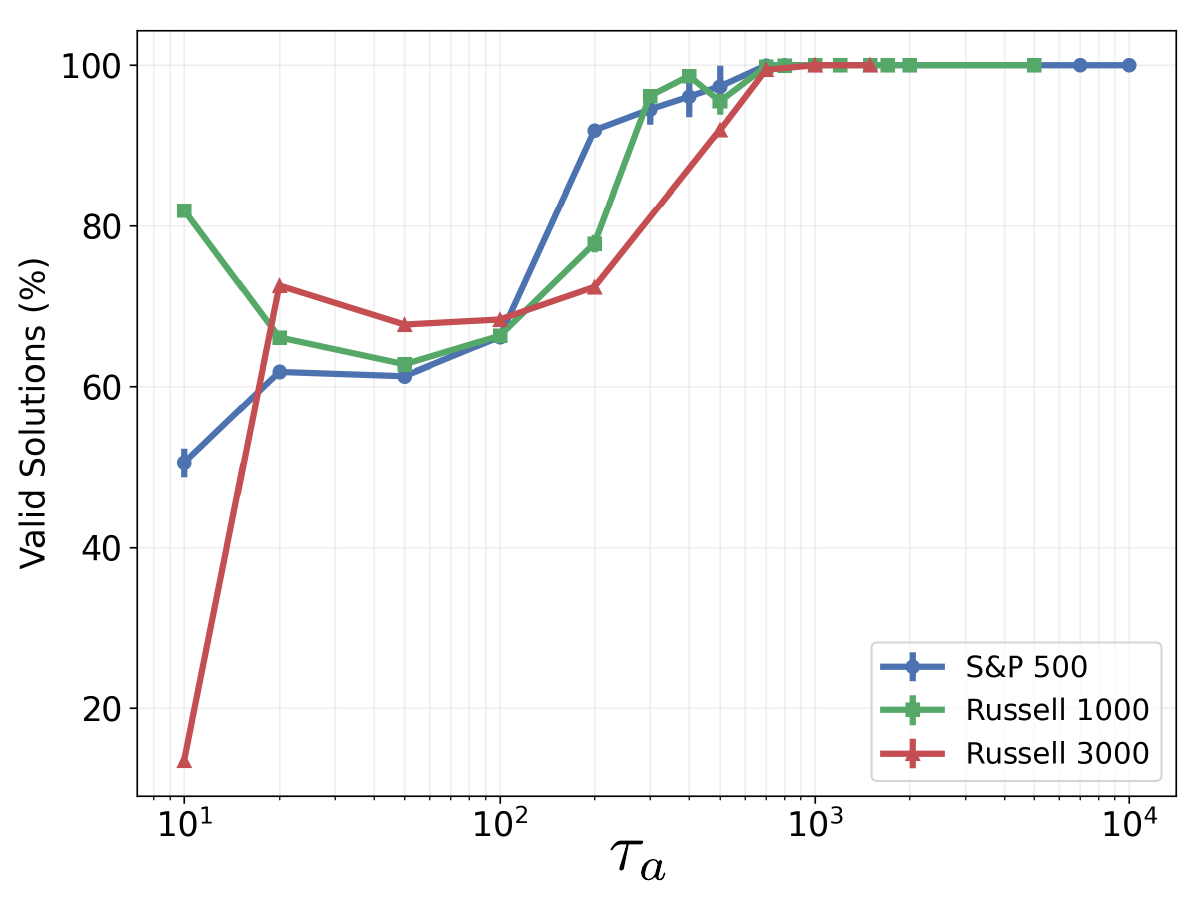}
    \caption{Percentage of valid solutions of VNA simulations with respect to the number of annealing steps $\tau_{a}$. The feasible solutions are extracted from the simulations presented in FIG.~\ref{fig:collapse}.}
    \label{fig:vna_valid_solutions_annealing}
\end{figure}

FIG.~\ref{fig:vna_valid_solutions_annealing} proposes a potential remedy. For simplicity, we consider only the normalization constraint and incorporate a Lagrangian multiplier term to increase the likelihood of its satisfaction (see the Hamiltonian in Eq.~(\ref{eq:criticality})). The figure shows the percentage of valid solutions for the S\&P 500, Russell 1000, and Russell 3000 as a function of the number of annealing steps, based on the same data used in FIG.~\ref{fig:collapse}. Averages are computed over five distinct trading months to mitigate correlations across objective function instances.

We observe that VNA achieves over 50\% feasibility after only $\tau_a = 50$ annealing steps, and reaches 100\% validity at longer annealing times for all three indices. These results suggest that incorporating Lagrangian terms can significantly enhance constraint satisfaction. However, identifying optimal combinations of Lagrangian multipliers remains challenging. Moreover, practical applications require that constraint satisfaction be enforced adaptively, since the full set of constraints may not be known a priori in real-world portfolio optimization tasks.

\section{Data Collapse and errors}\label{subsec:collapse_error}

The data collapse technique that we have implemented closely follows that introduced in \cite{Bhattacharjee_2001}. We start by concatenating, rescaling, and standardizing the raw data.

\begin{align}
        x_{i,j} &= \text{log}(\tau_a^iN_j^{\mu})\\
        y_{i,j} &= \text{log}(y_iN_j^{\kappa})\\
        Y_{i,j} &= \frac{y_{i,j}-\langle y_{i,j}\rangle}{\sqrt{\langle (y_{i,j}-\langle y_{i,j}\rangle)^2 \rangle}}.
\end{align}

Here, the indices $i$ and $j$ correspond to the annealing time and the number of assets, respectively. The logarithm of the raw data is taken for two main reasons: first, it linearizes the power law, which is easier for the spline to approximate; and second, it compresses the vast range on the y-axis into $O(1)$ numbers, making the optimization stable. Furthermore, the y-axis is standardized to have zero mean and unit variance; this stabilizes the optimizer and prevents it from driving the cost to zero by uniformly shifting or shrinking the y-axis. We now define the residual cost function as 

\begin{equation}
    \text{cost}_{\kappa,\mu} = \sum_{i,j} [Y_{i,j} - g_{\kappa,\mu}(x_{i,j})]^2,
\end{equation}

where, $g_{\kappa,\mu}$ is a cubic univariate spline used to approximate the universal scaling behavior. The cost function $\text{cost}_{\kappa,\mu}$ attains a minimum value of zero in the ideal case where the data corresponding to different system sizes $\{N_j\}$ collapse perfectly onto a single universal scaling function at the true value of $\kappa$ and $\mu$. With a finite number of data, we numerically minimize the cost function to extract the best estimates for the critical exponents: $\kappa^*$ and $\mu^*$.\\

The uncertainties in the critical exponents $\kappa$ and $\mu$ are obtained with a parametric bootstrap. To this end, we generate $N_{\text{boot}}$ number of synthetic datasets by adding independent Gaussian noise to the best-fit spline: $y^{\text{rep}}_{i,j} = g_{\kappa^*,\mu^*}(x_{i,j}) + \epsilon_{i,j}$ where, $\epsilon_{i,j} \sim \mathcal{N}(0,\sigma_{i,j}^2)$ ($\sigma_{i,j}$ is the standard error of the raw data). Each synthetic replica is processed through the same fitting pipeline as the original raw data to calculate a new estimate of the critical exponents, $\{\kappa^k,\mu^k\}$. The standard deviation of the resulting distribution $\{\kappa^k,\mu^k\}_{k=1}^{N_{\text{boot}}}$ is taken as the 1-$\sigma$ standard error of the corresponding critical exponents.

\bibliography{bibliography}

\end{document}